\documentclass[12pt]{article}
\pdfoutput=1

\usepackage[pdftex]{graphicx}
\usepackage{amssymb,amsmath,slashed}
\usepackage{cite}

\usepackage{longtable}
\usepackage{booktabs}
\usepackage{array}
\usepackage{ulem}
\usepackage{graphicx,epsfig}
\usepackage{amsmath}
\usepackage{tikz-feynman}
\usepackage{xcolor}
\definecolor{tit}{rgb}{0.1,0.2,0.4}

\renewcommand{\arraystretch}{1.2}

\newcommand{\eq}[1]{\begin{equation} #1 \end{equation}}

\newcommand{\eqa}[1]{\begin{eqnarray} #1 \end{eqnarray}}

\newcommand{\av}[1]{\langle #1 \rangle}

\newcommand{\MeV}{\,{\rm MeV}}
\newcommand{\GeV}{\,{\rm GeV}}
\newcommand{\TeV}{\,{\rm TeV}}
\newcommand{\heff}{\mathcal{H}_{\rm eff}}
\newcommand{\Aeff}{\mathcal{A}_{\rm eff}}

\newcommand{\Eq}[1]{Eq.~(\ref{#1})}
\newcommand{\Fig}[1]{Fig.~\ref{#1}}
\newcommand{\Sec}[1]{Section~\ref{#1}}
\newcommand{\App}[1]{App.~\ref{#1}}
\newcommand{\Reff}[1]{Ref.~\cite{#1}}
\newcommand{\Tab}[1]{Table~\ref{#1}}

\newcommand{\cD}{{\cal D}}
\newcommand{\cC}{{\cal C}}
\newcommand{\cO}{{\cal O}}
\newcommand{\cA}{{\cal A}}
\newcommand{\cR}{{\cal R}}

\newcommand{\Lag}{{\cal L}}
\newcommand{\comm}[2]{\left[#1,#2\right]}

\begin{document}

\begin{flushright}
CERN-TH-2021-140,
PSI-PR-21-22,
ZU-TH 46/21
\end{flushright}

$\ $
\vspace{1cm}
\begin{center}
\fontsize{16}{20}\selectfont
\bf 
Next-to-Leading-Order QCD Matching for $\Delta F=2$\\[2mm]
Processes in Scalar Leptoquark Models
\end{center}

\vspace{2mm}
\begin{center}
{\rm Andreas Crivellin{$^{\, a,b,c}$}, Jordi Folch Eguren{$^{\, d,e}$} and
Javier Virto{$^{\, d}$}}\\[5mm]
{\it\small
{$^{\, a}$} 
Physik-Institut, Universit\"at Z\"urich,
	Winterthurerstrasse 190, CH-8057 Z\"urich, Switzerland\\[2mm]
{$^{\, b}$} Paul Scherrer Institut, CH--5232 Villigen PSI, Switzerland\\[2mm]
{$^{\, c}$} CERN Theory Division, CH--1211 Geneva 23, Switzerland\\[2mm]
{$^{\, d}$} Departament de Física Quàntica i Astrofísica, Institut de Ciències del Cosmos,\\
Universitat de Barcelona, Martí Franquès 1, E08028 Barcelona, Catalunya\\[2mm]
{$^{\, e}$}Fakult\"at f\"ur Physik, TU Dortmund,
D-44221 Dortmund, Germany\\}
\end{center}

\vspace{1mm}
\begin{abstract}\noindent
Leptoquarks provide viable solutions to the flavour anomalies, i.e. they can explain the tensions between the measurements and the Standard Model predictions of the anomalous magnetic moment of the muon as well as $b\to s\ell^+\ell^-$ and $b\to c\tau\nu$ processes. However, LQs also contribute to other flavour observables, such as $\Delta F=2$ processes, at the loop-level. In particular, $B_s-\bar B_s$ mixing provides a crucial bound in setups addressing $b\to c\tau\nu$ data, often excluding a big portion of the parameter space that could otherwise account for it. In this article, we first derive the complete leading order matching, including all five scalar LQ representations, for $D^0-\bar D^0$, $K^0-\bar K^0$ and $B_{s,d}-\bar B_{s,d}$ mixing (at the dimension-six level). We then calculate the next-to-leading order $\alpha_s$ matching corrections to these $\Delta F=2$ processes in generic scalar leptoquark models. We find that the two-loop corrections increase the effects in $\Delta F=2$ processes by $\sim 5$-$10\%$ and significantly reduce the matching scale uncertainty.
\end{abstract}
\newpage
\setcounter{tocdepth}{2}
\tableofcontents

\allowdisplaybreaks

\section{Introduction}
\label{intro}
Leptoquarks (LQs) are hypothetical beyond the Standard Model (BSM) particles, arising originally in the context of Grand Unified Theories~\cite{Pati:1974yy,Georgi:1974sy,Georgi:1974yf,Fritzsch:1974nn}. What makes them special, and defines them, are their direct (common) couplings to leptons and quarks (i.e. they convert a quark into a lepton and vice versa). LQs were first systematically classified in Ref.~\cite{Buchmuller:1986zs}, where ten possible LQ representations under the Standard Model (SM) gauge group were found, of which five are scalar fields (spin 0) and five are vector (spin 1) particles. 

While LQs have received varying degrees of attention in the past, they have undergone a renaissance in recent years. This can be mainly attributed to the emergence of the flavour anomalies, i.e. the deviations from the SM predictions observed in several flavour observables. In particular,  $R(D^{(*)})$~\cite{Lees:2012xj,Lees:2013uzd,Aaij:2015yra,Aaij:2017deq,Aaij:2017uff,Abdesselam:2019dgh}, $b \rightarrow s \ell^+ \ell^-$ observables~\cite{CMS:2014xfa,Aaij:2015oid,Abdesselam:2016llu,Aaij:2017vbb,Aaij:2019wad,Aaij:2020nrf,Aaij:2021vac} and the muon anomalous magnetic moment $a_\mu$~\cite{Bennett:2006fi,Abi:2021gix} deviate from their SM predictions by more than $3\sigma$~\cite{Amhis:2016xyh,Murgui:2019czp,Shi:2019gxi,Blanke:2019qrx,Kumbhakar:2019avh}, $5\sigma$~\cite{Descotes-Genon:2015uva,Capdevila:2017bsm,Altmannshofer:2017yso,Alguero:2019ptt,Alok:2019ufo,Ciuchini:2019usw,Aebischer:2019mlg,Arbey:2019duh,Kumar:2019nfv,Alguero:2021anc,Altmannshofer:2021qrr,Alok:2020bia,Hurth:2020ehu,Ciuchini:2020gvn} and $4.2\sigma$~\cite{Aoyama:2020ynm,Aoyama:2012wk,Aoyama:2019ryr,Czarnecki:2002nt,Gnendiger:2013pva,Davier:2017zfy,Keshavarzi:2018mgv,Colangelo:2018mtw,Hoferichter:2019gzf,Davier:2019can,Keshavarzi:2019abf,Kurz:2014wya,Melnikov:2003xd,Masjuan:2017tvw,Colangelo:2017fiz,Hoferichter:2018kwz,Gerardin:2019vio,Bijnens:2019ghy,Colangelo:2019uex,Blum:2019ugy,Colangelo:2014qya}, respectively. 

It has been shown LQ models can account for $b\to s\ell^+\ell^-$ data~\cite{Gripaios:2014tna,Alonso:2015sja, Calibbi:2015kma, Hiller:2016kry, Bhattacharya:2016mcc, Buttazzo:2017ixm, Barbieri:2015yvd, Barbieri:2016las, Calibbi:2017qbu, Crivellin:2017dsk,DiLuzio:2017vat, Bordone:2018nbg, Kumar:2018kmr, Crivellin:2018yvo, Crivellin:2019szf, Cornella:2019hct, Bordone:2019uzc, Bernigaud:2019bfy,Aebischer:2018acj,Fuentes-Martin:2019ign,Popov:2019tyc,Fajfer:2015ycq,Blanke:2018sro,deMedeirosVarzielas:2019lgb,Varzielas:2015iva,Crivellin:2019dwb,Saad:2020ihm,Saad:2020ucl,Gherardi:2020qhc,DaRold:2020bib,Davighi:2020qqa,Iguro:2021kdw}, $R(D^{(*)})$~\cite{Alonso:2015sja, Calibbi:2015kma, Fajfer:2015ycq, Bhattacharya:2016mcc, Buttazzo:2017ixm, Barbieri:2015yvd, Barbieri:2016las, Calibbi:2017qbu, Bordone:2017bld, Bordone:2018nbg, Kumar:2018kmr, Biswas:2018snp, Crivellin:2018yvo, Blanke:2018sro, Heeck:2018ntp,deMedeirosVarzielas:2019lgb, Cornella:2019hct, Bordone:2019uzc,Sahoo:2015wya, Chen:2016dip, Dey:2017ede, Becirevic:2017jtw, Chauhan:2017ndd, Becirevic:2018afm, Popov:2019tyc,Fajfer:2012jt, Deshpande:2012rr, Freytsis:2015qca, Bauer:2015knc, Li:2016vvp, Zhu:2016xdg, Popov:2016fzr, Deshpand:2016cpw, Becirevic:2016oho, Cai:2017wry, Altmannshofer:2017poe, Kamali:2018fhr, Mandal:2018kau, Azatov:2018knx, Wei:2018vmk, Angelescu:2018tyl, Kim:2018oih, Aydemir:2019ynb, Crivellin:2019qnh, Yan:2019hpm,Crivellin:2017zlb, Marzocca:2018wcf, Bigaran:2019bqv,Crivellin:2019dwb,Saad:2020ihm,Dev:2020qet,Saad:2020ucl,Altmannshofer:2020axr,Fuentes-Martin:2020bnh,Gherardi:2020qhc,DaRold:2020bib,Iguro:2020cpg,Iguro:2018vqb} and/or $a_\mu$~\cite{Bauer:2015knc,Djouadi:1989md, Chakraverty:2001yg,Cheung:2001ip,Popov:2016fzr,Chen:2016dip,Biggio:2016wyy,Davidson:1993qk,Couture:1995he,Mahanta:2001yc,Queiroz:2014pra,ColuccioLeskow:2016dox,Chen:2017hir,Das:2016vkr,Crivellin:2017zlb,Cai:2017wry,Crivellin:2018qmi,Kowalska:2018ulj,Dorsner:2019itg,Crivellin:2019dwb,DelleRose:2020qak,Saad:2020ihm,Bigaran:2020jil,Dorsner:2020aaz,Fuentes-Martin:2020bnh,Gherardi:2020qhc,Babu:2020hun,Crivellin:2020tsz,DiLuzio:2018zxy}, making them prime candidates in the search for BSM models\footnote{The CMS excess in $pp\to e^+e^-$~\cite{CMS:2021ctt,Crivellin:2021rbf} could also be explained by LQs~\cite{Crivellin:2021egp,Crivellin:2021bkd}. The same is true of the Cabbibo Angle anomaly~\cite{Belfatto:2019swo,Grossman:2019bzp,Seng:2020wjq,Coutinho:2019aiy,Coutinho:2020xhc,Crivellin:2020lzu}, although some fine tuning in $D^0-\bar D^0$ mixing is required~\cite{Crivellin:2021egp}. Furthermore, scalar LQs are the only possible candidates for explaining $\Delta A_{FB}$~\cite{Carvunis:2021dss} in $b\to D^*\ell\nu$~\cite{Bobeth:2021lya}.}. As such, they have been studied in direct searches at the LHC~\cite{Kramer:1997hh,Kramer:2004df,Faroughy:2016osc,Greljo:2017vvb, Blumlein:1996qp, Dorsner:2017ufx, Cerri:2018ypt, Bandyopadhyay:2018syt, Hiller:2018wbv,Faber:2018afz,Schmaltz:2018nls,Chandak:2019iwj,Allanach:2019zfr, Buonocore:2020erb,Haisch:2020xjd,Borschensky:2020hot,Iguro:2020keo,Bandyopadhyay:2021pld,Bandyopadhyay:2020wfv}, leptonic observables~\cite{Crivellin:2020mjs} and oblique electroweak parameters, Higgs couplings to gauge bosons~\cite{Keith:1997fv,Dorsner:2016wpm,Bhaskar:2020kdr,Zhang:2019jwp,Gherardi:2020det,Crivellin:2020ukd}, a wide range of low energy precision probes~\cite{Shanker:1981mj,Shanker:1982nd,Leurer:1993em,Leurer:1993qx,Davidson:1993qk,Grossman:2019bzp,Seng:2020wjq,Belfatto:2019swo,Coutinho:2019aiy,Arnan:2019olv,Crivellin:2020lzu,Capdevila:2020rrl,Crivellin:2020ebi,Kirk:2020wdk,Alok:2020jod,Crivellin:2020oup,Crivellin:2020klg,Crivellin:2021njn,Belfatto:2021jhf,Branco:2021vhs,Bobeth:2017ecx,Dorsner:2019vgp,Mandal:2019gff,Crivellin:2021egp}. The complete scalar LQ Lagrangian and the corresponding set of Feynman rules has been presented recently in~\Reff{Crivellin:2021tmz}. 
The QCD corrections to LQ production and decay at colliders have been known for a long time~\cite{Plehn:1997az,Kramer:1997hh,Kramer:2004df} and have been improved to include NLO parton shower~\cite{Mandal:2015lca} or a large width~\cite{Hammett:2015sea}. Such QCD corrections have also been included in recent analyses correlating the $B$ anomalies to LHC searches~\cite{Faroughy:2016osc,Dorsner:2017ufx,Dorsner:2018ynv,Hiller:2018wbv,Monteux:2018ufc,Schmaltz:2018nls}. However, the calculation for the analogous $\alpha_s$ corrections to flavour observables is still incomplete. So far, only the $\cO(\alpha_s)$ corrections to semi-leptonic processes~\cite{Aebischer:2018acj} and $\ell\to\ell^\prime\gamma$~\cite{DelleRose:2020qak} have been calculated, but the analogous two-loop matching for $\Delta F=2$ processes is still missing. Here, specially in models aiming at an explanation of $R(D^{(*)})$, but also in models accounting for $b\to s\ell^+\ell^-$ (most importantly in models involving couplings to left-handed fermions only~\cite{DiLuzio:2017fdq,Crivellin:2019dwb}), $B_s-\bar B_s$ mixing provides a crucial constraint that limits the possible size of the new physics contribution.

In this article we calculate the next-to-leading order (NLO) QCD matching for $\Delta F=2$ processes in scalar LQ models.
However, we will first compute the one-loop matching for $D^0-\bar D^0$, $K^0-\bar K^0$ and $B_{s,d}-\bar B_{s,d}$ mixing, taking into account all five scalar LQ representations, which has so far not been presented in the literature. We then compute the two-loop $\cO(\alpha_s)$ corrections to these $\Delta F=2$ processes, which, together with the (known) two-loop QCD evolution of the corresponding effective operators~\cite{Ciuchini:1997bw,Buras:2000if}, is needed to reduce significantly the matching-scale uncertainty.
This is particularly important in light of the increasingly tighter constraints placed by $\Delta F=2$ processes on NP~\cite{Bona:2007vi,DiLuzio:2019jyq,Aebischer:2020dsw}.

\section{Scalar Leptoquark models at Low Energies}

\subsection{Leptoquark interactions with SM fermions}

\begin{table}
\centering
\begin{tabular}{l | c c r  } & $SU(3)$& {$SU(2)_L$}&$U(1)_Y$\\
\hline
$\Phi_1$& 3&1&$-{1}/{3}\ $\\ 
$\tilde{\Phi}_1$& 3 &1&$-{4}/{3}\ $\\
$\Phi_2$&3&2&${7}/{6}\ $\\
$\tilde{\Phi}_2$&3&2&${1}/{6}\ $\\
$\Phi_3$&3&3&$-{1}/{3}\ $
\end{tabular}
\caption{The five possible scalar leptoquark representations under the SM gauge group.}
\label{tab:Reps}
\end{table}

There are five different representations under the SM gauge group for scalar particles (i.e. scalar LQs)  such that a common coupling to quarks and leptons is possible~\cite{Buchmuller:1986zs}, as given in~\Tab{tab:Reps}\footnote{Here we disregard couplings to two quarks, which would lead to proton decay, and can be forbidden by assigning lepton and baryon numbers to the LQs.}.
The corresponding Lagrangian can be written as
\eqa{
\Lag &=& \left( {\lambda _{fj}^{1R}{\mkern 1mu} \bar u_f^c{\ell _j} + \lambda _{fj}^{1L}{\mkern 1mu} \bar Q_f^{{\kern 1pt} c}i{\tau _2}{L_j}} \right)\Phi _1^\dag  + \tilde \lambda _{fj}^1{\mkern 1mu} \bar d_f^c{\ell _j}\tilde \Phi _1^\dag  + \lambda _{fj}^{2RL}{\mkern 1mu} {{\bar u}_f}\Phi _2^Ti{\tau _2}{L_j}
\nonumber\\
&&+ \lambda _{fj}^{2LR}{\mkern 1mu} {{\bar Q}_f}{\ell _j}{\Phi _2} + \tilde \lambda _{fj}^2{\mkern 1mu} {{\bar d}_f}\tilde \Phi _2^Ti{\tau _2}{L_j} + \lambda _{fj}^3{\mkern 1mu} \bar Q_f^{{\kern 1pt} c}i{\tau _2}{\left( {\tau \cdot{\Phi _3}} \right)^\dag }{L_j} + {\rm{h}}{\rm{.c}}{\rm{.}}\ .
\label{LagLQ}
}
Here $Q$ and $L$ are the quark and lepton $SU(2)_L$ doublets while $u$, $d$ and $\ell$ are singlets, $\tau_2$ is the second Pauli matrix and the superscript $c$ denotes charge conjugation. 

After electroweak (EW) symmetry breaking, the $SU(2)_L$ doublets are decomposed into their components with definite electric charge and the quark and lepton fields can be transformed to the physical mass eigenbasis. While the rotations of the lepton and the right-handed quark fields can be absorbed by a redefinition of the couplings $\lambda$, and are thus unphysical, the CKM matrix unavoidably appears in couplings involving left-handed quark fields. Working in the down basis, such that the CKM matrix only enters in couplings to left-handed up-quarks, 
\eqa{
{\Lag_{{\rm{EW}}}} &=& 
\left( {\lambda _{fj}^{1R} \bar u_f^c P_R{\ell _j} + V_{ff'}^*\lambda _{f'j}^{1L} \bar u_f^cP_L{\ell _j} - \lambda _{fj}^{1L}{\mkern 1mu} \bar d_f^cP_L{\nu _j}} \right)\Phi _1^{ - 1/3*} 
+ \tilde \lambda _{fj}^1{\mkern 1mu} \bar d_f^cP_R{\ell _j}\tilde \Phi _1^{ - 4/3*}
\nonumber\\
&&  + \lambda _{fj}^{2RL} \left( {{\bar u}_f}P_L{\ell _j}{\Phi _2^{5/3} - {{\bar u}_f}P_L{\nu _j}}\Phi _2^{2/3} \right) 
+  {{V_{ff'}}\lambda _{f'j}^{2LR}{\mkern 1mu} {{\bar u}_f}P_R{\ell _j}\Phi _2^{5/3} + \lambda _{fj}^{2LR}{\mkern 1mu} {{\bar d}_f}P_R{\ell _j}\Phi _2^{2/3}}
\nonumber\\
&&  +\tilde \lambda _{fj}^2 \left( {{\bar d}_f}P_L{\ell _j}\tilde \Phi _2^{2/3} - {{\bar d}_f}P_L{\nu _j} \tilde \Phi _2^{ - 1/3}\right) 
+ V_{ff'}^*\lambda _{f'j}^3 \left( {\sqrt 2 {{\bar u}^c}P_L\nu \Phi _3^{2/3*} - {{\bar u}^c}{P_L}\ell \Phi _3^{ - 1/3*}} \right)
\nonumber\\
&& - \lambda _{fj}^3 \left( {{\bar d}^c}{P_L}{\nu \Phi _3^{ - 1/3*} + \sqrt 2 {{\bar d}^c}{P_L}\ell \Phi _3^{ - 4/3*}} \right) + \text{h.c.}\ ,
}
where the superscripts on the LQ fields indicate the corresponding electric charge.

We can write these interaction terms in a generic manner~\footnote{
The lepton $\ell$ here may be a charged lepton of a neutrino.
For $\Phi_1$, $\Phi_3$, $\ell$ denotes a charge-conjugated lepton field. This is inconsequential in our calculation of QCD corrections, since QCD does not see the charge of the lepton fields.  
}
\eq{
\Lag_{q\ell}^\text{LQ} = \sum_a \bar q\left( {{\Gamma^{aL}_{q\ell}}{P_L} + {\Gamma^{aR}_{ q\ell}}{P_R}} \right)\ell\, {\Phi_a} + {\rm{h}}{\rm{.c}}{\rm{.}}\ ,
\label{LQgen}
}
with $a=1,2,\dots$ running over the different LQ fields (several `copies' of LQs belonging to the same representation also allowed). In this notation we would have e.g. for $\Phi_a = \tilde \Phi_2^{2/3}$ the relation
\eq{
\Gamma^{aL}_{q\ell} = \Gamma^{aR}_{q\ell} = \tilde\lambda^2_{q\ell}\ ,
\label{LagLQgeneric}
}
and similarly for the other LQs. Thus, we will present our results in terms of the generic couplings $\Gamma^{a(L,R)}_{q\ell}$, and one can derive from them the specific cases in terms of the couplings in~\Eq{LagLQ}.

\subsection{Leptoquark contributions to $\Delta F = 2$ processes at LO}
\label{LOresults}

Due to the relatively low energy scale at which neutral meson mixing takes place, its physics can be described by an effective field theory (EFT) where SM EW-scale particles ($W, Z$, higgs and top) as well as the LQs\,\footnote{
We assume that the LQs are heavier than the EW scale as indicated by LHC searches.} are not dynamical degrees of freedom but are rather integrated out from the action. Flavour-changing transitions are then mediated by effective operators of dimension six or higher (see, e.g.~\cite{Aebischer:2017gaw}). The most general set of (physical) dimension six operators for $\Delta F=2$ processes contains eight operators (for a specific flavour transition)
\eq{
\Lag_\text{eff}^{\Delta F=2} =
- \sum_{i=1}^5 C_i \cO_i - \sum_{i=1}^3 C'_i \cO'_i \ .
}
In the case of $B_s-\bar B_s$ mixing, the operators in the so-called ``SUSY" basis read explicitly
\begin{align}
{\cO_1} &= {{\bar s}_\alpha }{\gamma ^\mu }{P_L}{b_\alpha }{{\bar s}_\beta }{\gamma _\mu }{P_L}{b_\beta }\ ,
&
{\cO_5} &= {{\bar s}_\alpha }{P_L}{b_\beta}{{\bar s}_\beta }{P_R}{b_\alpha}\ ,
\nonumber\\
{\cO_2} &= {{\bar s}_\alpha }{P_L}{b_\alpha}{{\bar s}_\beta }{P_L}{b_\beta}\ ,
&
{\cO'_1} &= {{\bar s}_\alpha }{\gamma ^\mu }{P_R}{b_\alpha}{{\bar s}_\beta }{\gamma _\mu }{P_R}{b_\beta}\ ,
\nonumber\\
{\cO_3} &= {{\bar s}_\alpha }{P_L}{b_\beta}{{\bar s}_\beta }{P_L}{b_\alpha}\ ,
&
{\cO'_2} &= {{\bar s}_\alpha }{P_R}{b_\alpha}{{\bar s}_\beta }{P_R}{b_\beta}\ ,
\label{operators}
\\
{\cO_4} &= {{\bar s}_\alpha }{P_L}{b_\alpha}{{\bar s}_\beta }{P_R}{b_\beta}\ ,
&
{\cO'_3} &= {{\bar s}_\alpha }{P_R}{b_\beta}{{\bar s}_\beta }{P_R}{b_\alpha}\ .
\nonumber
\end{align}
where $\alpha,\beta$ are colour indices. The corresponding expressions for $K^0-\bar K^0$, $D^0-\bar D^0$ and $B_d-\bar B_d$ mixing follow by a simple exchange of flavours.
The associated Wilson coefficients $C_i^{(\prime)}$ can be calculated from a given UV-complete theory by performing a matching calculation at the matching scale $\mu_0$, which is of the order of the mass of the particles that are integrated out. The matching calculation is done by equating the (expanded) full-theory and the EFT amplitudes at the matching scale, order by order in perturbation theory. We therefore write
\eq{
C_i = C_i^{(0)} + C_i^{(1)} + \cO(\alpha_s^2) \ ,
}
where the superscript indicates the order in the strong coupling $\alpha_s$.

\begin{figure}
	\begin{minipage}[t]{.5\textwidth}
		\begin{center}
			\begin{tikzpicture}
			\begin{feynman}
		    \vertex (a);
		    \vertex [below left=of a] (i1) {\(s\)};
		    \vertex [      right=of a ] (b);
		    \vertex [      above=of b ] (c);
		    \vertex [      left =of c ] (d);
		    \vertex [      below=of d ] (a);
		    \vertex [below right=of b ] (i2) {\(b\)};
		    \vertex [above right=of c ] (f2) {\(s\)};
		    \vertex [above  left=of d ] (f1) {\(b\)};
		    \diagram* {
		    	(i1) -- [anti fermion] (a) -- [fermion, edge label'=\(\ell'\)] (b) -- [scalar, edge label'=\(\Phi_b\)] (c) -- [fermion, edge label'=\(\ell\)] (d) -- [scalar, edge label'=\(\Phi_a\)] (a),
		    	(b) -- [anti fermion] (i2),
		    	(c) -- [fermion] (f2),
		    	(d) -- [anti fermion] (f1),
		    };
			\end{feynman}
			\end{tikzpicture}
		\end{center}
	\end{minipage}\hfill
	\begin{minipage}[t]{.5\textwidth}
		\begin{center}
			\begin{tikzpicture}
			\begin{feynman}
			\vertex (a);
			\vertex [below left=of a] (i1) {\(s\)};
			\vertex [      right=of a ] (b);
			\vertex [      above=of b ] (c);
			\vertex [      left =of c ] (d);
			\vertex [      below=of d ] (a);
			\vertex [below right=of b ] (i2) {\(b\)};
			\vertex [above right=of c ] (f2) {\(s\)};
			\vertex [above  left=of d ] (f1) {\(b\)};
			\diagram* {
				(i1) -- [anti fermion] (a) -- [scalar, edge label'=\(\Phi_b\)] (b) -- [fermion, edge label'=\(\ell'\)] (c) -- [scalar, edge label'=\(\Phi_a\)] (d) -- [fermion, edge label'=\(\ell\)] (a),
				(b) -- [anti fermion] (i2),
				(c) -- [fermion] (f2),
				(d) -- [anti fermion] (f1),
			};
			\end{feynman}
			\end{tikzpicture}
		\end{center}
	\end{minipage}
	\caption{Feynman diagrams depicting leading-order scalar LQ contributions to the  Wilson coefficients $C_i^{(0)}$.}
	\label{LOdiagrams}
\end{figure}
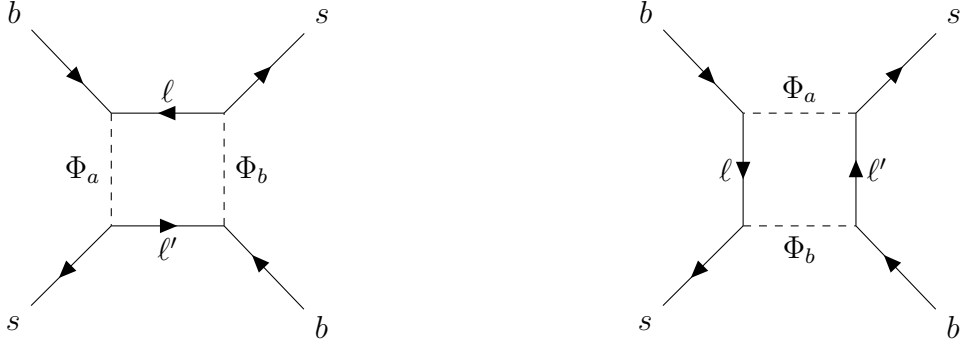

The leading-order contributions originating from LQ exchange to the LO Wilson coefficients $C_i^{(0)}$ arise from the one-loop diagrams shown in~\Fig{LOdiagrams}. These results, which are similar to the ones in the MSSM~\cite{Ciuchini:1998ix,Ciuchini:2007cw,Virto:2009wm,Altmannshofer:2009ne,Crivellin:2009ar,Crivellin:2010ys,Virto:2011yx,Mescia:2012fg} or 2HDMs~\cite{Crivellin:2013wna} are known~\cite{Arnan:2016cpy,Bobeth:2017ecx,Arnan:2019uhr,Aebischer:2020dsw}, and the non-zero Wilson coefficients at the matching scale $\mu_0$ are given by
\eqa{
\label{LQC1}
C_1^{(0)}(\mu_0) &=&  \frac1{{128{\pi ^2}{M^2}}}
\sum_{a,b} \Gamma _{ab}^R\Gamma _{ba}^R \ \frac{\log(x_a/x_b)}{x_a-x_b} 
\ ,\\
C_1^{\prime(0)}(\mu_0) &=&  \frac1{{128{\pi ^2}{M^2}}}
\sum_{a,b} \Gamma _{ab}^L\Gamma _{ba}^L \ \frac{\log(x_a/x_b)}{x_a-x_b}  
\ ,\\
\label{LQC4}
C_4^{(0)}(\mu_0) &=& -\frac1{{32{\pi ^2}{M^2}}}
\sum_{a,b} \Gamma _{ab}^L\Gamma _{ba}^R \ \frac{\log(x_a/x_b)}{x_a - x_b}
\ ,
}
with
\eq{
x_a = \frac{M_a^2}{M^2}\ ,
\qquad
\Gamma _{ab}^X = \sum_\ell  \Gamma _{s\ell }^{aX}\Gamma _{b\ell }^{bX*}\ .
}
In these expressions we have introduced a generic mass $M$ setting scale for the Wilson coefficients. In the degenerate case where $M_a=M$ (for all $a$)
\eq{
C_1^{(0)}(\mu_0)=\frac{1}{128\pi^2M^2}\sum_a\Gamma_{aa}^R\Gamma_{aa}^R\ ,
\qquad
C^{\prime(0)}_1 = C_1\big|_{R\to L}\ ,
\qquad
C_4^{(0)}(\mu_0)=\frac{-1}{32\pi^2M^2}\sum_a\Gamma_{aa}^L\Gamma_{aa}^R\ .
\label{LOdegenerate}
}
Note that the leading-order Wilson coefficients $C_i^{(0)}$ do not depend explicitly on the matching scale~$\mu_0$. However they do carry an implicit dependence 
through the scale dependence of the LQ masses $M_a$ and the couplings~$\Gamma_{ab}^X$. For the numerical analysis it will be reasonable to define the parameters inside $C_i^{(0)}(\mu_0)$ as the renormalized parameters at the scale $\mu_0$, i.e.  $M_a(\mu_0)$ and $\Gamma_{ab}^X(\mu_0)$. Formally, setting a different scale amounts to an $\alpha_s$ correction to $C_i^{(0)}$, and thus one could absorb the $\mu_0$ dependence into $C_i^{(1)}$. However, since the renormalization-scale dependence of $M_a$ and  $\Gamma_{ab}^X$ is known prior to $C_i^{(1)}$ (which we are calculating in this article), it is reasonable to include this implicit matching scale dependence in $C_i^{(0)}$ already at leading order.

\bigskip

Below the matching scale, the renormalization-scale dependence of the Wilson coefficients is determined by the anomalous dimension matrix (ADM) in the EFT
\eq{
\frac{dC_i}{d\log\mu} = \gamma_{ji}\, C_j =
\Big(\hat\alpha_s \gamma_{ji}^{(0)} + \hat\alpha_s^2 \gamma_{ji}^{(1)} + \cdots \Big)\, C_j\,,
\label{rge}
}
with $\hat\alpha_s \equiv \alpha_s/(4\pi)$.
The leading-order ADM $\gamma^{(0)}$ is obtained from the one-loop renormalization of the EFT, and it is scheme-independent~\footnote{
Here and in the following all elements in the 2-3 sector will appear {\color{gray}in gray} in order to make it clear that they do not play any role when $C_{2,3}=0$, as is our case.
}:
\eq{
\label{ADMLO}
\gamma^{(0)} = \left(
\begin{array}{ccccc}
4 & 0 & 0 & 0 & 0 \\
0 & {\color{gray}-28/3} & {\color{gray}4/3} & 0 & 0 \\
0 & {\color{gray}16/3} & {\color{gray}32/3} & 0 & 0 \\
0 & 0 & 0 & -16 & 0  \\
0 & 0 & 0 & -6 & 2
\end{array}
\right)
\ ,
}
for $C_{1-5}$, with the ADMs for $C'_{1-3}$ equal to the ones for the $C_{1-3}$ sector.

The next-to-leading ADM $\gamma^{(1)}$ arises at two-loops~\cite{Buras:2000if,Ciuchini:1997bw}. It was derived in~\Reff{Buras:2000if} using a different basis for physical operators (the ``BMU" basis).
The Wilson coefficients at NLO will be scheme-dependent, and it will be important to use the same scheme for $\gamma^{(1)}$ in order to get scheme-independent observables. In our basis and scheme, the ADM $\gamma^{(1)}$ is given by (see~\App{app:basis} for details)
\eq{
\gamma^{(1)}=
\left(
\begin{array}{ccccc}
\frac{4 f}{9}-7 & 0 & 0 & 0 & 0 \\
0 & {\color{gray}\frac{220 f}{27}-\frac{476}{3}} & {\color{gray}-\frac{4 f}{27}-\frac{4}{3}} & 0 & 0 \\
0 & {\color{gray}73+\frac{110 f}{27}} & {\color{gray}\frac{359}{3}-\frac{218 f}{27}} & 0 & 0 \\
0 & 0 & 0 & \frac{68 f}{9}-\frac{1343}{6} & 4 f-\frac{225}{2} \\
0 & 0 & 0 & \frac{22 f}{3}-99 & \frac{71}{3}-\frac{22 f}{9} \\
\end{array}
\right)\ ,
\label{gamma1}
}
where $f$ is the number of active quark flavours.

\subsection{LO matching for $\Delta F=2$ processes including $SU(2)$ invariance}

We now calculate the leading-order effect in $\Delta F=2$ processes taking into account explicitly $SU(2)$ invariance for the different LQ representations. For $D^0-\bar D^0$ mixing we find
\begin{equation}
\begin{aligned}
C_1^{D(0)}(\mu_0) &= \frac{{  1}}{{128{\pi ^2}{M^2}}}\sum\limits_{i,j}^{} {\left( \begin{array}{l}
	\hat \lambda _{1i}^{1L*}{\mkern 1mu} \hat \lambda _{2i}^{1L}{\mkern 1mu} \hat \lambda _{1j}^{1L*}{\mkern 1mu} \hat \lambda _{2j}^{1L}f\left( {{x_1},{x_1}} \right) + \lambda _{1i}^{2LR}\lambda _{2i}^{2LR*}\lambda _{1j}^{2LR}\lambda _{2j}^{2LR*}f\left( {{x_2},{x_2}} \right)\\
	+ 5\hat \lambda _{1i}^{3*}\hat \lambda _{2i}^3\hat \lambda _{1j}^{3*}\hat \lambda _{2j}^3f\left( {{x_3},{x_3}} \right) + 2\lambda _{1i}^{1L*}\hat \lambda _{2i}^3\hat \lambda _{1j}^{3*}\lambda _{2j}^{1L}f\left( {{x_1},{x_3}} \right){\mkern 1mu} {\mkern 1mu} 
	\end{array} \right)} \,,\\
C_4^{D(0)}(\mu_0) &= \frac{-1}{{32{\pi ^2}{M^2}}}\sum\limits_{i,j}^{} {\left( {\hat \lambda _{1i}^{1L*}\hat \lambda _{2i}^{1L}\lambda _{1j}^{1R*}\lambda _{2j}^{1R}f\left( {{x_1},{x_1}} \right) + \hat \lambda _{1i}^{2LR}\hat \lambda _{2i}^{2LR*}\hat \lambda _{1j}^{2RL}\hat \lambda _{2j}^{2RL*}f\left( {{x_2},{x_2}} \right)} \right)} \,,\\
C_1^{\prime D(0)}(\mu_0) &= \frac{{  1}}{{128{\pi ^2}{M^2}}}\sum\limits_{i,j}^{} {\left( {\lambda _{1i}^{1R*}\lambda _{2i}^{1R}\lambda _{1j}^{1R*}\lambda _{2j}^{1R}f\left( {{x_1},{x_1}} \right) + 2\lambda _{1i}^{2RL}\lambda _{2i}^{2RL*}\lambda _{1j}^{2RL}\lambda _{2j}^{2RL*}f\left( {{x_2},{x_2}} \right)} \right)}\,, 
\end{aligned}
\end{equation}
with
\eq{
f\left( {x,y} \right) = \frac{\log (x/y)}{x - y}\,,
}
and where we have defined
\eq{
\hat \lambda _{fj}^{1L}  = \sum\limits_{f' = 1}^{3} {V_{ff'}^*\lambda _{f'j}^{1L}}\ ,
\quad
\hat \lambda _{fj}^{2LR}  = \sum\limits_{f' = 1}^{3} {{V_{ff'}}\lambda _{f'j}^{2LR}} \ ,
\quad
\hat \lambda _{fj}^3  = \sum\limits_{f' = 1}^{3} {V_{ff'}^*\lambda _{f'j}^3} \ .
}
For Kaon mixing we find
\eqa{
C_1^{K (0)}(\mu_0)
&=& \frac{{  1}}{{128{\pi ^2}{M^2}}}\sum\limits_{i,j}^{} {\left( \begin{array}{l}
\lambda _{1i}^{1L*}{\mkern 1mu} \lambda _{2i}^{1L}{\mkern 1mu} \lambda _{1j}^{1L*}{\mkern 1mu} \lambda _{2j}^{1L}f\left( {{x_1},{x_1}} \right) + \lambda _{1i}^{2LR}\lambda _{2i}^{2LR*}\lambda _{1j}^{2LR}\lambda _{2j}^{2LR*}f\left( {{x_2},{x_2}} \right)\\
+ 5\lambda _{1i}^{3*}\lambda _{2i}^3\lambda _{1j}^{3*}\lambda _{2j}^3f\left( {{x_3},{x_3}} \right) + 2\lambda _{1i}^{1L*}\lambda _{2i}^3{\mkern 1mu} \lambda _{1j}^{3*}\lambda _{2j}^{1L}f\left( {{x_1},{x_3}} \right){\mkern 1mu} 
\end{array} \right)}\ ,
\nonumber\\
C_1^{\prime K(0)}(\mu_0)
&=& \frac{{  1}}{{128{\pi ^2}{M^2}}}\sum\limits_{i,j}^{} {\left( \begin{array}{l}
{\mkern 1mu} \tilde \lambda _{1i}^{1*}{\mkern 1mu} \tilde \lambda _{2i}^1{\mkern 1mu} \tilde \lambda _{1j}^{1*}{\mkern 1mu} \tilde \lambda _{2j}^1{\mkern 1mu} f\left( {{x_{\tilde 1}},{x_{\tilde 1}}} \right) + 2\tilde \lambda _{1i}^{2*}{\mkern 1mu} \tilde \lambda _{2i}^2{\mkern 1mu} \tilde \lambda _{1j}^{2*}{\mkern 1mu} \tilde \lambda _{2j}^2f\left( {{x_{\tilde 2}},{x_{\tilde 2}}} \right){\mkern 1mu} \\
+ 2\lambda _{1i}^{2LR}\tilde \lambda _{2i}^{2*}\tilde \lambda _{1j}^2\lambda _{2j}^{2LR*}f\left( {{x_2},{x_{\tilde 2}}} \right)
\end{array} \right)} \ .
}
From this formula, the matching conditions for $B_s-\bar B_s$ and $B_d-\bar B_d$ mixing can be obtained by a trivial exchange of flavour indices. 

Comparing these expressions with Eqs.~(\ref{LQC1})-(\ref{LQC4}) we can make the following identification between the couplings $\Gamma$ in the generic Lagrangian of~\Eq{LagLQgeneric} and the LQ couplings in the $SU(2)$-invariant Lagrangian of~\Eq{LagLQ}.
For the case of $D^0-\bar D^0$ mixing we have
\begin{equation}
\begin{array}{l}
\sum\limits_{a,b}^{} {\Gamma _{ab}^R\Gamma _{ba}^Rf\left( {{x_a},{x_b}} \right) = } \sum\limits_{i,j}^{} {\left( \begin{array}{l}
	\hat \lambda _{1i}^{1L*}{\mkern 1mu} \hat \lambda _{2i}^{1L}{\mkern 1mu} \hat \lambda _{1j}^{1L*}{\mkern 1mu} \hat \lambda _{2j}^{1L}f\left( {{x_1},{x_1}} \right) + \lambda _{1i}^{2LR}\lambda _{2i}^{2LR*}\lambda _{1j}^{2LR}\lambda _{2j}^{2LR*}f\left( {{x_2},{x_2}} \right)\\
	+ 5\hat \lambda _{1i}^{3*}\hat \lambda _{2i}^3\hat \lambda _{1j}^{3*}\hat \lambda _{2j}^3f\left( {{x_3},{x_3}} \right) + 2\lambda _{1i}^{1L*}\hat \lambda _{2i}^3\hat \lambda _{1j}^{3*}\lambda _{2j}^{1L}f\left( {{x_1},{x_3}} \right){\mkern 1mu} {\mkern 1mu} 
	\end{array} \right)} {\mkern 1mu} \,,\\
\sum\limits_{a,b}^{} {\Gamma _{ab}^L\Gamma _{ba}^Lf\left( {{x_a},{x_b}} \right) = \sum\limits_{i,j}^{} {\left( {\lambda _{1i}^{1R*}\lambda _{2i}^{1R}\lambda _{1j}^{1R*}\lambda _{2j}^{1R}f\left( {{x_1},{x_1}} \right) + 2\lambda _{1i}^{2RL}\lambda _{2i}^{2RL*}\lambda _{1j}^{2RL}\lambda _{2j}^{2RL*}f\left( {{x_2},{x_2}} \right)} \right)} } \,,\\
\sum\limits_{a,b}^{} {\Gamma _{ab}^L\Gamma _{ba}^Rf\left( {{x_a},{x_b}} \right)}  = \sum\limits_{i,j}^{} {\left( {\hat \lambda _{1i}^{1L*}\hat \lambda _{2i}^{1L}\lambda _{1j}^{1R*}\lambda _{2j}^{1R}f\left( {{x_1},{x_1}} \right) + \hat \lambda _{1i}^{2LR}\hat \lambda _{2i}^{2LR*}\hat \lambda _{1j}^{2RL}\hat \lambda _{2j}^{2RL*}f\left( {{x_2},{x_2}} \right)} \right)} \,,
\end{array}\end{equation}
while for $K^0-\bar K^0$ mixing we have
\begin{equation}
\begin{array}{l}
\sum\limits_{a,b}^{} {\Gamma _{ab}^R\Gamma _{ba}^Rf\left( {{x_a},{x_b}} \right) = } \sum\limits_{i,j}^{} {\left( \begin{array}{l}
	\lambda _{1i}^{1L*}{\mkern 1mu} \lambda _{2i}^{1L}{\mkern 1mu} \lambda _{1j}^{1L*}{\mkern 1mu} \lambda _{2j}^{1L}f\left( {{x_1},{x_1}} \right) + \lambda _{1i}^{2LR}\lambda _{2i}^{2LR*}\lambda _{1j}^{2LR}\lambda _{2j}^{2LR*}f\left( {{x_2},{x_2}} \right)\\
	+ 5\lambda _{1i}^{3*}\lambda _{2i}^3\lambda _{1j}^{3*}\lambda _{2j}^3f\left( {{x_3},{x_3}} \right) + 2\lambda _{1i}^{1L*}\lambda _{2i}^3{\mkern 1mu} \lambda _{1j}^{3*}\lambda _{2j}^{1L}f\left( {{x_1},{x_3}} \right){\mkern 1mu} 
	\end{array} \right)} {\mkern 1mu} \,,\\
\sum\limits_{a,b}^{} {\Gamma _{ab}^L\Gamma _{ba}^Lf\left( {{x_a},{x_b}} \right) = \sum\limits_{i,j}^{} {\left( \begin{array}{l}
		{\mkern 1mu} \tilde \lambda _{1i}^{1*}{\mkern 1mu} \tilde \lambda _{2i}^1{\mkern 1mu} \tilde \lambda _{1j}^{1*}{\mkern 1mu} \tilde \lambda _{2j}^1{\mkern 1mu} f\left( {{x_{\tilde 1}},{x_{\tilde 1}}} \right) + 2\tilde \lambda _{1i}^{2*}{\mkern 1mu} \tilde \lambda _{2i}^2{\mkern 1mu} \tilde \lambda _{1j}^{2*}{\mkern 1mu} \tilde \lambda _{2j}^2f\left( {{x_{\tilde 2}},{x_{\tilde 2}}} \right){\mkern 1mu} \\
		+ 2\lambda _{1i}^{2LR}\tilde \lambda _{2i}^{2*}\tilde \lambda _{1j}^2\lambda _{2j}^{2LR*}f\left( {{x_2},{x_{\tilde 2}}} \right)
		\end{array} \right)} } \,.
\end{array}
\end{equation}
These expressions can be used to write any Wilson coefficient calculated using the generic Lagrangian of~\Eq{LagLQgeneric} in terms of the couplings of the $SU(2)$-invariant Lagrangian, also beyond the LO.

\section{Next-to-Leading-Order Calculation}
\label{NLOcalc}

\subsection{QCD renormalization of the LQ Lagrangian}
\label{correctionsNLO}

\begin{figure}
	\begin{center}
		\begin{tabular}{cp{7mm}c}
			\includegraphics[width=0.7\textwidth]{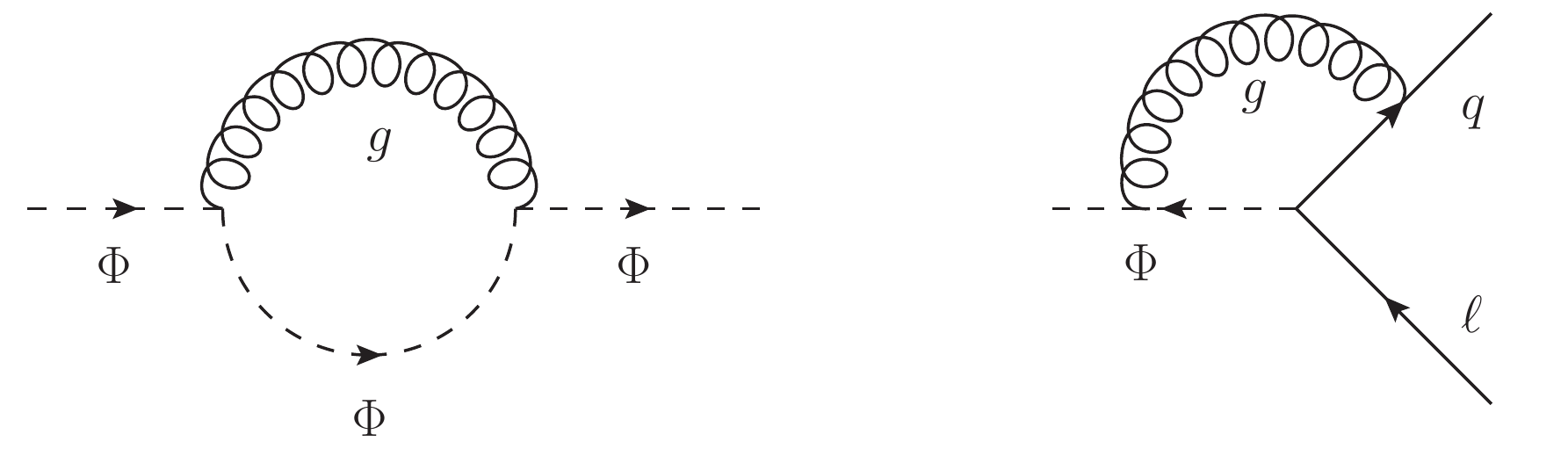}
		\end{tabular}
	\end{center}
	\caption{One-loop diagrams leading to the renormalization of the LQ mass and the LQ coupling to quarks and leptons.}
	\label{LQrenorm}
\end{figure}

In the presence of next-to-leading order (NLO) QCD corrections, the LQ Lagrangian must be renormalized. Thus, the couplings and fields in~\Eq{LagLQgeneric} are to be understood as bare (divergent) quantities. Using multiplicative renormalization, the Lagrangian reads
\eqa{
\Lag_\text{LQ} &=&-Z_{\Phi}\partial^{\mu}\Phi^{*}\partial_{\mu}\Phi-Z_MM^2\Phi^{*}\Phi
\nonumber\\
&&
-\frac{Z_G}{4}G^{A\mu\nu}G^{A}_{\mu\nu} +iZ_{g_s} g_s\left[\Phi^{*}_{\alpha}\partial^{\mu}\Phi_{\beta}-(\partial^{\mu}\Phi^{*}_{\alpha})\Phi_{\beta}\right)]
G^{A}_{\mu}T^{A}_{\alpha\beta}-Z_{g_s}^2 g_s^2\Phi^{*}_{\alpha}\Phi_{\beta} G^{A\mu}G^{B}_{\mu}T^{A}_{\alpha \sigma}T^{B}_{\sigma\beta}
\nonumber\\
&&
+Z_q\bar{q}_{\alpha}(i \slashed{\partial}) q_{\beta}\delta_{\alpha\beta}+Z_{g_s} g_s\bar{q}_{\alpha}\slashed{G}^{A}q_{\beta}T_{\alpha\beta}^{A}
+ \big[ Z_\Gamma \,\bar{q}\left( \Gamma_{q\ell}^{aL} P_L + \Gamma_{q\ell}^{aR} P_R \right)\ell\,\Phi_a^{*}+\text{h.c.} \big]
\ ,
}
where we have considered massless quarks, only a single LQ as well as only one generation of quark and leptons. However, as QCD is flavour blind, this trivially generalizes to the case of multiple generations of quarks and leptons as well as several LQ components. The renormalization constants $Z_i=1+\delta_i$ contain the counterterms $\delta_i$. At one loop order, these counterterms are fixed by subtracting the $1/\epsilon$ poles originating from the diagrams shown in~\Fig{LQrenorm} (as well as the quark self-energy) within the $\overline{\text{MS}}$ scheme, which we will use throughout this article, resulting in
\begin{align}
\begin{aligned}
\delta_M &=-\hat\alpha_sC_F\frac{1}{\epsilon}\ ,
\qquad\delta_{\Phi}=\hat\alpha_sC_F\frac{2}{\epsilon}\ ,
\\
\delta_\Gamma &=-\hat\alpha_sC_F\frac{1}{\epsilon}\ ,
\qquad \delta_q=-\hat\alpha_sC_F\frac{1}{\epsilon}\ ,
\\
\delta_{g_s} &=-\hat\alpha_s\left(-C_F-\frac{C_A}{2}\right)\frac{1}{\epsilon}\,.
\end{aligned}
\end{align}
The renormalized LQ mass and the couplings thus obey a renormalization group (RG) equation which determines their renormalization scale dependence:
\eqa{
\frac{dM_a}{d\log\mu} &=& - 3C_F \hat\alpha_s M_a + \cO(\hat\alpha_s^2)\ ,
\nonumber\\
\frac{d\Gamma _{q\ell }^{aX}}{d\log\mu} &=& - 3C_F \hat\alpha_s \Gamma _{q\ell }^{aX} + \cO(\hat\alpha_s^2)
\ ,
\label{betafcts}\\
\frac{d\hat\alpha_s}{d\log\mu} &=&  \cO(\hat\alpha_s^2)
\ ,
\nonumber
}
where $\hat\alpha_s \equiv \alpha_s/(4\pi)$ and $C_F=4/3$.

\subsection{Set-up of the NLO matching calculation}
\label{setup}

The set-up for the NLO matching calculation is the same as the one described e.g. in~\Reff{Virto:2009wm}. We match the amplitudes of the full theory onto the ones arising in the EFT at NLO in $\alpha_s$ at the matching scale $\mu_0$ to determine $C_i^{(1)}$. The amplitudes in the EFT are given by
\eq{
\Aeff = \sum_{i,j} C_i \, \Big(\delta_{ij} + \hat\alpha_s \,r_{ij} + \cO(\alpha_s^2) \Big) \av{\cO_j}^{(0)}\ ,
\label{EFTamp}
}
where $\av{\cO_i}^{(0)}$ are tree-level matrix elements, and $r_{ij}$ is given by~\cite{Virto:2009wm}
\eq{
r = \left(
\begin{array}{ccccc}
4/3 & 0 & 0 & 0 & 0 \\
0 & {\color{gray}44/3} & {\color{gray}-4/3} & 0 & 0 \\
0 & {\color{gray}-16/3} & {\color{gray}-16/3} & 0 & 0 \\
0 & 0 & 0 & 64/3 & 0  \\
0 & 0 & 0 & 6 & 10/3
\end{array}
\right) \log(\mu/\lambda)
+
\left(
\begin{array}{ccccc}
-5 & 0 & 0 & 0 & 0 \\
0 & {\color{gray}1/3} & {\color{gray}-1} & 0 & 0 \\
0 & {\color{gray}-15/2} & {\color{gray}-25/6} & 0 & 0 \\
0 & 0 & 0 & 19/3 & -3  \\
0 & 0 & 0 & -1/2 & -7/6
\end{array}
\right)
\ ,
\label{rmatrix}
}
with the elements in the $3\times 3$ primed sector equal to the ones in the $i=1,2,3$ sector.
The parameter $\lambda$, an artificial gluon mass, is an infrared (IR) regulator needed to separate ultraviolet (UV) and IR divergences. The $\lambda$ dependence must cancel in the matching procedure such that the Wilson coefficients do not depend on it. This also provides a cross-check of the calculation.
The log-independent term (the second matrix in~\Eq{rmatrix}) depends on the renormalization scheme. In this case the chosen scheme is the $\overline{\text{MS}}$-NDR scheme with the evanescent operators given in Appendix~A of~\Reff{Buras:2000if}.
It is important to use the same scheme in the calculation of the NLO full-theory amplitude in order to get consistent results.

The amplitudes in the full theory at NLO are the sum of the LO contributions from the diagrams in~\Fig{LOdiagrams} and the NLO contributions from the two-loop diagrams shown in~\Fig{NLOdiagrams}. It can be written as
\eq{
\cA_\text{LQ} = \sum_{i} \Big(F_i^{(0)} + \hat\alpha_s \,F_i^{(1)}  + \cO(\alpha_s^2) \Big) \av{\cO_i}^{(0)}\ ,
\label{LQamp}
}
again in terms of tree-level matrix elements. Requiring equality of EFT and full-theory amplitudes at the matching scale $\mu_0$ order-by-order in $\alpha_s(\mu_0)$, and writing
\eq{
C_i(\mu_0) = C_i^{(0)}(\mu_0) + C_i^{(1)}(\mu_0) + \cO(\alpha_s(\mu_0)^2)\ ,
}
gives
\eqa{
C_i^{(0)} &=& F_i^{(0)}\ ,
\\
C_i^{(1)} &=& \hat\alpha_s\, F_i^{(1)} - \hat\alpha_s \sum_j F_j^{(0)} r_{ji}\ .
}
The coefficients $C_i^{(0)} = F_i^{(0)}$ have been given in~\Sec{LOresults}. The only missing pieces are thus the NLO functions $F_i^{(1)}$, which are obtained from the evaluation of the genuine two-loop Feynman diagrams and one-loop diagrams with counterterms to be discussed in the next section.

\subsection{Calculation of the two-loop contributions}\label{32section}

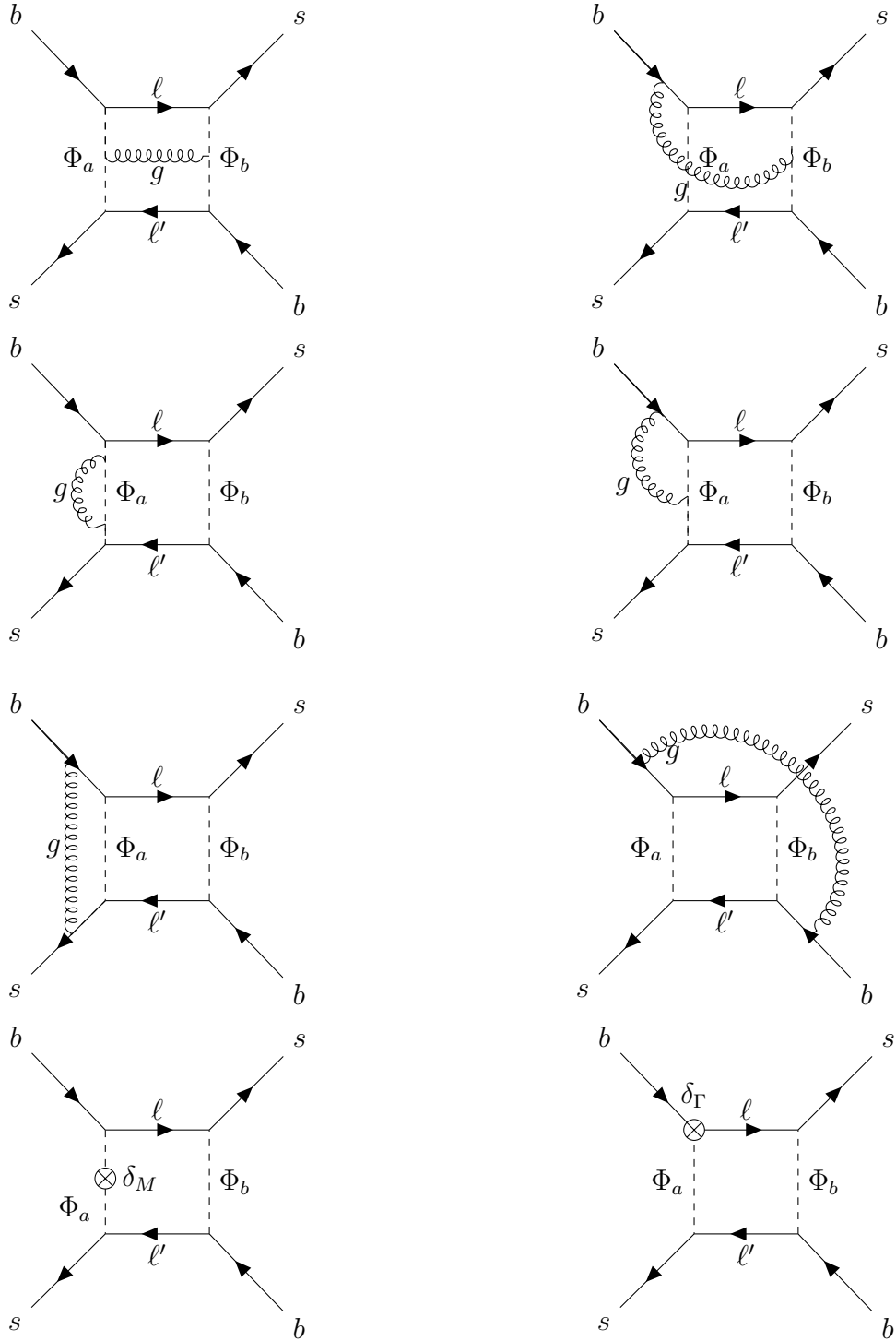
\begin{figure}
	\label{2loopdiagrams}
	\begin{minipage}[t]{.5\textwidth}
		\begin{center}
			\begin{tikzpicture}
			\begin{feynman}
			\vertex (a);
			\vertex [below left=of a] (i1) {\(s\)};
			\vertex [      right=of a ] (b);
			\vertex [      above=of b ] (c);
			\vertex [      left =of c ] (d);
			\vertex [      below=of d ] (a);
			\vertex [below right=of b ] (i2) {\(b\)};
			\vertex [above right=of c ] (f2) {\(s\)};
			\vertex [above  left=of d ] (f1) {\(b\)};
			\vertex [below=0.7cm of d] (l1);
			\vertex [below=0.7 cm of c] (l2);
			\diagram* {
				(i1) -- [anti fermion] (a) -- [anti fermion, edge label'=\(\ell'\)] (b) -- [scalar, edge label'=\(\Phi_b\)] (c) -- [anti fermion, edge label'=\(\ell\)] (d) -- [scalar, edge label'=\(\Phi_a\)] (a),
				(d) -- [scalar] (l1) -- [gluon, edge label'=\(g\)] (l2),
				(b) -- [anti fermion] (i2),
				(c) -- [fermion] (f2),
				(d) -- [anti fermion] (f1),
			};
			\end{feynman}
			\end{tikzpicture}
		\end{center}
	\end{minipage}\hfill
	\begin{minipage}[t]{.5\textwidth}
		\begin{center}
			\begin{tikzpicture}
			\begin{feynman}
			\vertex (a);
			\vertex [below left=of a] (i1) {\(s\)};
			\vertex [      right=of a ] (b);
			\vertex [      above=of b ] (c);
			\vertex [      left =of c ] (d);
			\vertex [below right=of b ] (i2) {\(b\)};
			\vertex [above right=of c ] (f2) {\(s\)};
			\vertex [above  left=of d ] (f1) {\(b\)};
			\vertex [above left=0.5cm of d] (l1);
			\vertex [above=0.9 cm of b] (l2);
			\diagram* {
				(i1) -- [anti fermion] (a) -- [anti fermion, edge label'=\(\ell'\)] (b) -- [scalar, edge label'=\(\Phi_b\)] (c) -- [anti fermion, edge label'=\(\ell\)] (d) -- [scalar, edge label=\(\Phi_a\)] (a),
				(f1) -- [plain] (l1) -- [gluon, edge label'=\(g\), half right] (l2),
				(b) -- [anti fermion] (i2),
				(c) -- [fermion] (f2),
				(d) -- [anti fermion] (f1),
			};
			\end{feynman}
			\end{tikzpicture}
		\end{center}
	\end{minipage}
	\begin{minipage}[t]{.5\textwidth}
		\begin{center}
			\begin{tikzpicture}
			\begin{feynman}
			\vertex (a);
			\vertex [below left=of a] (i1) {\(s\)};
			\vertex [      right=of a ] (b);
			\vertex [      above=of b ] (c);
			\vertex [      left =of c ] (d);
			\vertex [      below=of d ] (a);
			\vertex [below right=of b ] (i2) {\(b\)};
			\vertex [above right=of c ] (f2) {\(s\)};
			\vertex [above  left=of d ] (f1) {\(b\)};
			\vertex [below=0.3cm of d] (l1);
			\vertex [above=0.3 cm of a] (l2);
			\diagram* {
				(i1) -- [anti fermion] (a) -- [anti fermion, edge label'=\(\ell'\)] (b) -- [scalar, edge label'=\(\Phi_b\)] (c) -- [anti fermion, edge label'=\(\ell\)] (d) -- [scalar, edge label=\(\Phi_a\)] (a),
				(d) -- [scalar] (l1) -- [gluon, edge label'=\(g\), half right] (l2) -- [scalar] (a),
				(b) -- [anti fermion] (i2),
				(c) -- [fermion] (f2),
				(d) -- [anti fermion] (f1),
			};
			\end{feynman}
			\end{tikzpicture}
		\end{center}
	\end{minipage}\hfill
	\begin{minipage}[t]{.5\textwidth}
		\begin{center}
			\begin{tikzpicture}
			\begin{feynman}
			\vertex (a);
			\vertex [below left=of a] (i1) {\(s\)};
			\vertex [      right=of a ] (b);
			\vertex [      above=of b ] (c);
			\vertex [      left =of c ] (d);
			\vertex [below right=of b ] (i2) {\(b\)};
			\vertex [above right=of c ] (f2) {\(s\)};
			\vertex [above  left=of d ] (f1) {\(b\)};
			\vertex [above left=0.5cm of d] (l1);
			\vertex [above=0.7 cm of a] (l2);
			\diagram* {
				(i1) -- [anti fermion] (a) -- [anti fermion, edge label'=\(\ell'\)] (b) -- [scalar, edge label'=\(\Phi_b\)] (c) -- [anti fermion, edge label'=\(\ell\)] (d) -- [scalar, edge label=\(\Phi_a\)] (a),
				(f1) -- [plain] (l1) -- [gluon, edge label'=\(g\), half right] (l2) -- [scalar] (a),
				(b) -- [anti fermion] (i2),
				(c) -- [fermion] (f2),
				(d) -- [anti fermion] (f1),
			};
			\end{feynman}
			\end{tikzpicture}
		\end{center}
	\end{minipage}
	\begin{minipage}[t]{.5\textwidth}
		\begin{center}
			\begin{tikzpicture}
			\begin{feynman}
			\vertex (a);
			\vertex [below left=of a] (i1) {\(s\)};
			\vertex [      right=of a ] (b);
			\vertex [      above=of b ] (c);
			\vertex [      left =of c ] (d);
			\vertex [below right=of b ] (i2) {\(b\)};
			\vertex [above right=of c ] (f2) {\(s\)};
			\vertex [above  left=of d ] (f1) {\(b\)};
			\vertex [above left=0.7cm of d] (l1);
			\vertex [below left=0.7 cm of a] (l2);
			\diagram* {
				(i1) -- [anti fermion] (a) -- [anti fermion, edge label'=\(\ell'\)] (b) -- [scalar, edge label'=\(\Phi_b\)] (c) -- [anti fermion, edge label'=\(\ell\)] (d) -- [scalar, edge label=\(\Phi_a\)] (a),
				(f1) -- [plain] (l1) -- [gluon, edge label'=\(g\)] (l2) -- [scalar] (a),
				(b) -- [anti fermion] (i2),
				(c) -- [fermion] (f2),
				(d) -- [anti fermion] (f1),
			};
			\end{feynman}
			\end{tikzpicture}
		\end{center}
	\end{minipage}\hfill
	\begin{minipage}[t]{.5\textwidth}
		\begin{center}
			\begin{tikzpicture}
			\begin{feynman}
			\vertex (a);
			\vertex [below left=of a] (i1) {\(s\)};
			\vertex [      right=of a ] (b);
			\vertex [      above=of b ] (c);
			\vertex [      left =of c ] (d);
			\vertex [below right=of b ] (i2) {\(b\)};
			\vertex [above right=of c ] (f2) {\(s\)};
			\vertex [above  left=of d ] (f1) {\(b\)};
			\vertex [above left=0.7cm of d] (l1);
			\vertex [below right=0.7 cm of b] (l2);
			\diagram* {
				(i1) -- [anti fermion] (a) -- [anti fermion, edge label'=\(\ell'\)] (b) -- [scalar, edge label'=\(\Phi_b\)] (c) -- [anti fermion, edge label'=\(\ell\)] (d) -- [scalar, edge label'=\(\Phi_a\)] (a),
				(f1) -- [plain] (l1) -- [gluon, half left] (l2),
				(b) -- [anti fermion] (i2),
				(c) -- [fermion] (f2),
				(d) -- [anti fermion] (f1),
			};
			\vertex [above=0.8em of d] {\(g\)};
			\end{feynman}
			\end{tikzpicture}
		\end{center}
	\end{minipage}
	\begin{minipage}[t]{.5\textwidth}
		\begin{center}
			\begin{tikzpicture}
			\begin{feynman}
			\vertex (a);
			\vertex [below left=of a] (i1) {\(s\)};
			\vertex [      right=of a ] (b);
			\vertex [      above=of b ] (c);
			\vertex [      left =of c ] (d);
			\vertex [      below=of d ] (a);
			\vertex [below right=of b ] (i2) {\(b\)};
			\vertex [above right=of c ] (f2) {\(s\)};
			\vertex [above  left=of d ] (f1) {\(b\)};
			\diagram* {
				(i1) -- [anti fermion] (a) -- [anti fermion, edge label'=\(\ell'\)] (b) -- [scalar, edge label'=\(\Phi_b\)] (c) -- [anti fermion, edge label'=\(\ell\)] (d) -- [scalar] (a),
				(b) -- [anti fermion] (i2),
				(c) -- [fermion] (f2),
				(d) -- [anti fermion] (f1),
			};
			\vertex [crossed dot, fill=white] (i) at ($(a)!0.0!(b) - (0, -0.8)$) {};
			\vertex [below left=1.5em of i] {\(\Phi_a\)};
			\vertex [right=0.5cm of i] {\(\delta_M\)};
			\end{feynman}
			\end{tikzpicture}
		\end{center}
	\end{minipage}
	\begin{minipage}[t]{.5\textwidth}
		\begin{center}
			\begin{tikzpicture}
			\begin{feynman}
			\vertex (a);
			\vertex [below left=of a] (i1) {\(s\)};
			\vertex [      right=of a ] (b);
			\vertex [      above=of b ] (c);
			\vertex [      left =of c ] (d);
			\vertex [      below=of d ] (a);
			\vertex [below right=of b ] (i2) {\(b\)};
			\vertex [above right=of c ] (f2) {\(s\)};
			\vertex [above  left=of d ] (f1) {\(b\)};
			\diagram* {
				(i1) -- [anti fermion] (a) -- [anti fermion, edge label'=\(\ell'\)] (b) -- [scalar, edge label'=\(\Phi_b\)] (c) -- [anti fermion, edge label'=\(\ell\)] (d) -- [scalar, edge label'=\(\Phi_a\)] (a),
				(b) -- [anti fermion] (i2),
				(c) -- [fermion] (f2),
				(d) -- [anti fermion] (f1),
			};
			\vertex [crossed dot, fill=white] (i) at ($(a)!0.0!(b) - (0, -1.5)$) {};
			\vertex [above=0.5cm of i] {\(\delta_{\Gamma}\)};
			\end{feynman}
			\end{tikzpicture}
		\end{center}
	\end{minipage}
	\caption{A sample of two-loop diagrams, and one-loop diagrams with counterterm insertions (indicated by the cross), contributing to the NLO QCD matching for $B_s-\bar B_s$ mixing. }
	\label{NLOdiagrams}
\end{figure}

In order to extract the NLO functions $F_i^{(1)}$, we compute the $\cO(\alpha_s)$ part of the (renormalized) amplitude in the full theory  $\cA(b_\alpha \bar s_\beta\to s_\delta \bar b_\gamma)$, at vanishing external momenta.
We express this part of the amplitude as a sum of the two-loop Feynman diagrams ($\cD_x$) and one-loop Feynman diagrams with counterterms ($\cC_x$), shown in~\Fig{NLOdiagrams},
\eq{
i\cA_\text{LQ}^\text{NLO} = \sum_{x\in\{\text{NLO\ diagrams}\}} \big(\cD_x + \cC_x\big)\ .
}
The counterterm diagrams have the structure of a one-loop box diagram with a $1/\epsilon$ vertex or propagator insertion, and thus the corresponding one-loop integral must be calculated up to and including terms of order $\epsilon$.
The pairs $(\cD_x + \cC_x)$ are UV-finite, and can be written as
\eq{
\cD_x + \cC_x = \hat\alpha_s f_x(m_j,\lambda)
(\bar u_{s\delta} \Gamma_x u_{b\alpha})
(\bar v_{s\beta} \tilde\Gamma_x v_{b\gamma})
+
\hat\alpha_s f'_x(m_j,\lambda)
(\bar u_{s\delta} \Gamma'_x v_{b\gamma})
(\bar v_{s\beta} \tilde\Gamma'_x u_{b\alpha})\ .
\label{Di+Ci}
}
The parameter $\lambda$ is a gluon mass that we have introduced to regularize IR divergencies in diagrams where the gluon connects the external (massless) quark legs. This is the same IR regulator appearing in~\Eq{rmatrix}.

In order to extract the functions $F_i^{(1)}$ we must write the spinor structures in~\Eq{Di+Ci} as a linear combination of tree-level matrix elements $\av{\cO_i}^{(0)}$. We do this by applying suitable Dirac projectors $P_i = P_A^{(i)}S^{(i)}\otimes P_B^{(i)}S^{(i)}$ 
which act on the spinor structures as
\eq{
v_{b\gamma}\bar{u}_{s\delta}\rightarrow P_{A}^{(k)} S^{(k)}_{\delta\gamma}\ ,
\quad
u_{b\alpha}\bar{v}_{s\beta}\rightarrow P_{B}^{(k)} S^{(k)}_{\alpha\beta} \ .
}
Here $P^{(k)}$ are Dirac matrices and $S^{(k)}$ colour structures ($\delta$ or $T^A$). These projectors are defined such that
\eq{
P_i \av{\cO_j}^{(0)} = \delta_{ij} + \cO(\epsilon^2)\ .
}
Specific details about these projections are given in~\App{app:projectors}.
In this way, the contribution to the function $F_i^{(1)}$ from the pair $(\cD_x + \cC_x)$ is given by
\eq{
\alpha_s F_i^{(1)} \Big|_x = P_i[\cD_x + \cC_x]\ .
}
The advantage of this approach is that the projection can be performed before evaluating the loop integrals, transforming the integrands into scalar functions of the loop momenta.
The scalar two-loop integrals can now be computed as in~\Reff{Virto:2009wm}: First, loop momenta in the numerators are reduced by expressing them in the form of the denominators; second, the denominators are decomposed using partial fraction, after which the integral can be expressed as a sum of terms of the form
\eq{
f(M_i,\lambda) \int \frac{d^dq_1\,d^dq_2}{(2\pi)^{2d}}
\frac1{(q_1^2 - m_1^2)^{n_1} (q_2^2 - m_2^2)^{n_2} [(q_1-q_2)^2 - m_3^2]^{n_3}}\ .
\label{MasterIntegrals}
}
The solution of these scalar integrals is known~\cite{Davydychev:1992mt}.

The contributions from each pair $(\cD_x + \cC_x)$ to the the functions $F_i^{(1)}$ are separately UV-finite, and this provides a non-trivial check of the two-loop integrals (note that the individual expressions for the scalar integrals in~\Eq{MasterIntegrals} contain $1/\epsilon$ and $1/\epsilon^2$ poles).
The results for the functions $F_i^{(1)}$ still depend on the IR regulator $\lambda$. This dependence is cancelled in the combination $F_i^{(1)} - F_j^{(0)} r_{ji}$. This cancellation is also non-trivial and constitutes yet another check of the two-loop calculation.

All types of two-loop diagrams are shown in ~\Fig{NLOdiagrams}. It is useful to classify these diagrams into finite, UV divergent and IR divergent ones. The first two diagrams are finite and thus no renormalization is required. The following two diagrams are UV divergent, and correspond precisely to the one-loop renormalization of the LQ self-energy and vertex correction from ~\Fig{LQrenorm}. Their corresponding counterterm diagrams are shown in the last row of~\Fig{NLOdiagrams}. The remaining diagrams, in which the gluon connects external quarks, are IR divergent. Such diagrams will have a $\lambda$ dependence which will contribute to the function $F_i^{(1)}$. As mentioned before, this $\lambda$ dependence will cancel with the one from the EFT contained in~\Eq{rmatrix} when performing the matching.

\subsection{Matching results for the Wilson coefficients at NLO}

The final results for the (non-zero) NLO Wilson Coefficients $C_i^{(1)}$ at the matching scale are
\begin{eqnarray}
C_1^{\left( 1 \right)}(\mu_0) &=&
 \dfrac{{{\alpha _s}\Gamma _{ab}^R\Gamma _{ba}^R}}{{4608{\pi ^3}{M^2}x_a^2x_b^2{{\left( {{x_a} - {x_b}} \right)}^2}}}
\nonumber\\
&&
\hspace{-8mm}
\times \left[ {x_a^3x_b^2\left( {30\text{Li}_2\left( {1 - X_b^a} \right) + \log \left( {X_b^a} \right)\left( {54\log \left( {X_b^\mu } \right) + 69} \right) - 12{{\log }^2}\left( {X_b^a} \right) - 5{\pi ^2} + 36} \right)} \right.
\nonumber\\
&&
\hspace{-8mm}
- x_a^4{x_b}\left( {6\text{Li}_2\left( {1 - X_b^a} \right) + 3{{\log }^2}\left( {X_b^a} \right) + 36\log \left( {X_b^a} \right) + {\pi ^2} + 36} \right)
\nonumber\\
&&
\hspace{-8mm}
\left. + 6x_a^5\left( {6\text{Li}_2\left( {1 - X_b^a} \right) + 3{{\log }^2}\left( {X_b^a} \right) + {\pi ^2}} \right) + a \leftrightarrow b \right]\ ,
\label{NLOC1}
\\
C_4^{\left( 1 \right)}(\mu_0) &=&
-\dfrac{{{\alpha _s}\Gamma _{ab}^L\Gamma _{ba}^R}}{{1152{\pi ^3}{M^2}{x_a}{x_b}{{\left( {{x_a} - {x_b}} \right)}^2}}}
\nonumber\\
&&
\hspace{-8mm}
\times \left[ {x_a^2{x_b}\left( {12\text{Li}_2\left( {1 - X_b^a} \right) + 6\log \left( {X_b^a} \right)\left( {4\log \left( {X_b^a} \right) - 3\left( {\log \left( {X_a^\mu } \right) + \log \left( {X_b^\mu } \right)} \right) + 10} \right) + {\pi ^2}} \right)} \right.
\nonumber\\
&&
\hspace{-8mm}
\left. { - x_b^3\left( {{\pi ^2} - 6\text{Li}_2\left( {1 - X_b^a} \right)} \right) + a \leftrightarrow b} \right]\ ,
\\
C_5^{\left( 1 \right)}(\mu_0) &=&
-\dfrac{{{\alpha _s}\Gamma _{ab}^L\Gamma _{ba}^R}}{{384{\pi ^3}{M^2}{x_a}{x_b}{{\left( {{x_a} - {x_b}} \right)}^2}}}
\label{NLOC5}
\\
&&
\hspace{-8mm}
\times \left[ {x_b^3\left( {{\pi ^2} - 6\text{Li}_2\left( {1 - X_b^a} \right)} \right) + {x_a}x_b^2\left( { 12\text{Li}_2\left( {1 - X_b^a} \right) - 12\log \left( {X_b^a} \right) - {\pi ^2}} \right) + a \leftrightarrow b} \right]\ ,
\nonumber
\end{eqnarray}
with 
\begin{eqnarray}
X^a_b=\dfrac{M_a^2}{M_b^2}\ ,
\qquad
X^\mu_a=\dfrac{\mu_0^2}{M_a^2}\ ,
\qquad
x_a=\dfrac{M_a^2}{M^2}\ .
\end{eqnarray}
Note that the dependence on $M$ drops out. In the equal LQ mass limit one has 
\begin{eqnarray}
C_1^{(1)}(\mu_0) &=& \alpha_s\frac{\Gamma^R_{aa}\Gamma^R_{aa}}{4608\pi^3M^2}\left(108 \log\frac{\mu_0}{M}+34\pi^2-273\right)\ ,
\label{NLOdegenerate1}
\\[2mm]
C_4^{(1)}(\mu_0) &=& \alpha_s\frac{\Gamma^L_{aa}\Gamma^R_{aa}}{1152\pi^3M^2}\left(72 \log\frac{\mu_0}{M}+2\pi^2-105\right)\ ,
\\[2mm]
C_5^{(1)}(\mu_0)&=& -\alpha_s\frac{\Gamma^L_{aa}\Gamma^R_{aa}}{384\pi^3M^2}\left(2\pi^2+3\right)\ .
\label{NLOdegenerate5}
\end{eqnarray}
The result for $C^{\prime(1)}_1(\mu_0)$ is equal to that of $C^{(1)}_1(\mu_0)$ with the replacement $\Gamma_{ab}^R \to \Gamma_{ab}^L$.

The results derived here can be easily translated into a matching to the SMEFT above the EW scale. For the necessary formulas we refer to e.g.~\Reff{Aebischer:2020dsw}.

\section{Phenomenological analysis}
\label{Pheno}

\subsection{Numerical Results}

Let us now derive simple numerical results from the analytic expressions obtained in the previous section for $K^0-\bar K^0$, $D^0-\bar D^0$ and $B_{s,d}-\bar B_{s,d}$ mixing as a function of the couplings $\Gamma$ (for $\mu_0=M$).

\begin{table}
\centering
\setlength{\tabcolsep}{16pt}
\renewcommand{\arraystretch}{1.5}
\begin{tabular}{@{}ccccc@{}}
\toprule
& $K^0-\bar K^0$ & $D^0-\bar D^0$ & $B_d-\bar B_d$ & $B_s-\bar B_s$
\\
\midrule
$B^{(1)}_{P}(\mu)$ & $0.506(17)$ &  $0.757(27)$ & $0.913(86)$ & $0.952(66)$
\\
$B^{(2)}_{P}(\mu)$ & $0.46(3)$ &  $0.65(4)$ & $0.761(76)$ & $0.806(59)$
\\
$B^{(3)}_{P}(\mu)$ & $0.79(5)$ &  $0.96(8)$ & $1.07(22)$ & $1.10(16)$
\\
$B^{(4)}_{P}(\mu)$ & $0.78(5)$ & $0.87(6)$ & $1.040(87)$ & $1.022(66)$
\\
$B^{(5)}_{P}(\mu)$ & $0.47(4)$ & $0.68(5)$ & $0.96(10)$ & $0.943(75)$
\\
\bottomrule
\end{tabular}
\caption{Bag parameters calculated within lattice QCD, adapted from Refs.~\cite{Carrasco:2015pra,FermilabLattice:2016ipl}. The renormalization scale is $\mu = \{ 3, 3, 4.18, 4.18 \}\GeV$  for $P=\{ K^0, D^0, B_d, B_s \}$. }
\label{tab:bagpars}
\end{table}

The relevant quantity is the matrix element of the $\Delta F=2$ effective Hamiltonian,
\eq{
\av{P^0| \heff^{\Delta F=2} |\bar P^0} = \sum_i C_i(\mu) \av{\cO_i(\mu)}\ ,
\label{eq:mixingamplitude}
}
where  $\av{\cO_i(\mu)}\equiv \av{P^0| \cO_i(\mu) |\bar P^0}$ can be expressed in terms of non-perturbative ``bag parameters'' $B^{(i)}_P$~(see e.g.~\Reff{FermilabLattice:2016ipl}),
\eqa{
\av{\cO_1^{(\prime)}(\mu)} &=& \frac23 f_{P}^2 M_{P}^2 B^{(1)}_{P}(\mu)\ ,
\label{MM1}
\\
\av{\cO_2^{(\prime)}(\mu)} &=& -\frac5{12} \bigg( \frac{M_{P}}{m_h(\mu)+m_l(\mu)} \bigg)^2\ f_{P}^2 M_{P}^2 B^{(2)}_{P}(\mu) \ ,
\\
\av{\cO_3^{(\prime)}(\mu)} &=& \frac1{12} \bigg( \frac{M_{P}}{m_h(\mu)+m_l(\mu)} \bigg)^2\ f_{P}^2 M_{P}^2 B^{(3)}_{P}(\mu) \ ,
\\
\av{\cO_4(\mu)} &=& \frac12 \bigg[ \bigg( \frac{M_{P}}{m_h(\mu)+m_l(\mu)} \bigg)^2 + \frac16\bigg] \ f_{P}^2 M_{P}^2 B^{(4)}_{P}(\mu) \ ,
\\
\av{\cO_5(\mu)} &=& \frac16 \bigg[ \bigg( \frac{M_{P}}{m_h(\mu)+m_l(\mu)} \bigg)^2 + \frac32\bigg]\ f_{P}^2 M_{P}^2 B^{(5)}_{P}(\mu) \ ,
\label{MM5}
}
where $P=\{ K^0, D^0, B_d, B_s \}$ and $(m_h, m_l)=\{(m_s,m_d),(m_c,m_u),(m_b,m_d),(m_b,m_s)\}$ are running $\overline{\text{MS}}$ masses.
The numerical values for the bag parameters $B^{(i)}_{P}(\mu)$ are calculated using lattice QCD and can be found in Refs.~\cite{Carrasco:2015pra,FermilabLattice:2016ipl}\,\footnote{
Other recent determinations can be found in~Refs.\cite{DiLuzio:2019jyq,Bazavov:2017weg,Dowdall:2019bea}.
}. For convenience we reproduce these numbers in~\Tab{tab:bagpars} adjusted to the conventions used in~Eqs.~(\ref{MM1})-(\ref{MM5}). The quoted results for the bag parameters are given at the renormalization scales  $\mu = \{ 3, 3, 4.18, 4.18 \}\GeV$  for $P=\{ K^0, D^0, B_d, B_s \}$, and in the renormalization scheme of~\Reff{Buras:2000if}, which is the same one used here in the calculation of the Wilson coefficients. The numerical values of the various quantities appearing in ~Eqs.~(\ref{MM1})-(\ref{MM5}) are collected in~\Tab{tabInputs}. The resulting numbers for the matrix elements $\av{\cO_i(\mu)}$ at the relevant renormalization scales are collected in~\Tab{tab:matrixelements}.

\begin{table}
\centering
\setlength{\tabcolsep}{22pt}
\renewcommand{\arraystretch}{1.5}
\begin{tabular}{@{}ll@{}}
\toprule
$M_{K^0} = 497.611(13) \MeV$ \cite{PDG}&
$M_{D^0} = 1.86484(5) \GeV$ \cite{PDG} \\
$M_{B_d} = 5.27965(12) \GeV$  \cite{PDG}&
$M_{B_s} = 5.36688(14) \GeV$  \cite{PDG} \\
$\overline m_u(3\GeV) = 2.3(2) \MeV$ ${}^{\dagger\dagger}$ &
$\overline m_d(3\GeV) = 4.4(2) \MeV$ ${}^{\dagger\dagger}$ \\
$\overline m_s(3\GeV) = 84.4(6) \MeV$ ${}^{\dagger\dagger}$ &
$\overline m_c(3\GeV) = 0.988(7) \GeV$  \cite{FLAG} \\
$\overline m_b(\overline m_b) = 4.18(3) \GeV$ \cite{PDG} &
\\
$\overline m_d(\overline m_b) = 4.1(2) \MeV$ ${}^{\dagger\dagger}$ &
$\overline m_s(\overline m_b) = 78.9(6) \MeV$ ${}^{\dagger\dagger}$ \\
$f_{K} = 155.7(3) \MeV$ \cite{FLAG} &
$f_{D} = 212.0(7) \MeV$ \cite{FLAG}\\
$f_{B_d} = 190.0(1.3) \MeV$  \cite{FLAG} &
$f_{B_s} = 230.3(1.3) \MeV$   \cite{FLAG} \\
\bottomrule
\end{tabular}
\caption{Set of inputs used in the numerical analysis. The inputs marked ${}^{\dagger\dagger}$ have been obtained from the values at the scale of $2\GeV$ given in~\Reff{FLAG}, by running them to $3\GeV$ and $4.18\GeV$ using RunDec~\cite{Chetyrkin:2000yt} at four loops in 4-flavour QCD.
}
\label{tabInputs}
\end{table}

In order to provide numerical formulas for the matrix element in~\Eq{eq:mixingamplitude} we also need the matching result $C_i(\mu_0)$ and the evolution matrix $U(\mu,\mu_0)$, defined by
\eq{
C_i(\mu) = U(\mu,\mu_0)_{ij} C_j(\mu_0)\,.
}
The evolution matrix is calculated by solving the RGE in~\Eq{rge} numerically, using the LO and NLO ADMs in~Eqs.(\ref{ADMLO}) and~(\ref{gamma1}).
For the evolution of the strong coupling $\alpha_s(\mu)$ we use the four loop result from RunDec~\cite{Chetyrkin:2000yt}. We find
\eqa{
U(4.18\GeV,1\TeV) &=&
\left(
\begin{array}{ccccc}
0.794 & 0 & 0 & 0 & 0 \\
0 & {\color{gray}1.886} & {\color{gray}-0.392} & 0 & 0 \\
0 & {\color{gray}-0.079} & {\color{gray}0.520} & 0 & 0 \\
0 & 0 & 0 & 2.909 & 0.666 \\
0 & 0 & 0 & 0.114 & 0.902
\end{array}
\right)\,,
\\[3mm]
U(3\GeV,1\TeV) &=&
\left(
\begin{array}{ccccc}
0.775 & 0 & 0 & 0 & 0 \\
0 & {\color{gray}2.034} & {\color{gray}-0.445} & 0 & 0 \\
0 & {\color{gray}-0.089} & {\color{gray}0.484} & 0 & 0 \\
0 & 0 & 0 & 3.299 & 0.798 \\
0 & 0 & 0 & 0.148 & 0.898
\end{array}
\right)\,.
}
Note that the evolution for $C'_{1,2,3}$ is the same as the one for $C_{1,2,3}$. 

For the LQ contribution to the Wilson coefficients at the matching scale $C_i(\mu_0)=C_i(1\TeV)$ we use the formulas in~Eqs.(\ref{LOdegenerate}) and (\ref{NLOdegenerate1})-(\ref{NLOdegenerate5}) with $M=\mu_0=1\TeV$ (the matching scale dependence will be discussed in the following section).
We find:
\eqa{
C_1(1\TeV) &=&
8.30 \cdot 10^{-10}\,\Gamma_{aa}^R\Gamma_{aa}^R\, \GeV^{-2}\ ,
\\
C'_1(1\TeV) &=& 
8.30 \cdot 10^{-10}\,\Gamma_{aa}^L\Gamma_{aa}^L\, \GeV^{-2}\ ,
\\
C_4(1\TeV) &=& 
-3.38 \cdot 10^{-9}\,\Gamma_{aa}^L\Gamma_{aa}^R\,\GeV^{-2}\ ,
\\
C_5(1\TeV) &=& 
-1.69 \cdot 10^{-10}\,\Gamma_{aa}^L\Gamma_{aa}^R\,\GeV^{-2}\ . 
}
Putting everything together, we have
\eqa{
\av{K^0| \heff^{\Delta F=2}|\bar K^0} &=&
\Big[
(0.131\pm 0.004)(\Gamma_{aa}^L\Gamma_{aa}^L + \Gamma_{aa}^R\Gamma_{aa}^R)
+ (-84.5\pm 5.5) \Gamma_{aa}^L\Gamma_{aa}^R
\Big] \cdot 10^{-11} \GeV^2\ ,
\nonumber
\\
\av{D^0| \heff^{\Delta F=2}|\bar D^0} &=&
\Big[
(0.051\pm 0.002)(\Gamma_{aa}^L\Gamma_{aa}^L + \Gamma_{aa}^R\Gamma_{aa}^R)
+ (-2.91\pm 0.20) \Gamma_{aa}^L\Gamma_{aa}^R
\Big] \cdot 10^{-9} \GeV^2\ ,
\nonumber
\\
\av{B_d| \heff^{\Delta F=2}|\bar B_d} &=&
\Big[
(0.41\pm 0.38)(\Gamma_{aa}^L\Gamma_{aa}^L + \Gamma_{aa}^R\Gamma_{aa}^R)
+ (-9.41\pm 0.81) \Gamma_{aa}^L\Gamma_{aa}^R
\Big] \cdot 10^{-9} \GeV^2\ ,
\nonumber
\\
\av{B_s| \heff^{\Delta F=2}|\bar B_s} &=&
\Big[
(0.64\pm 0.04)(\Gamma_{aa}^L\Gamma_{aa}^L + \Gamma_{aa}^R\Gamma_{aa}^R)
+ (-13.98\pm 0.94) \Gamma_{aa}^L\Gamma_{aa}^R
\Big] \cdot 10^{-9} \GeV^2\ .
\nonumber
\\
}
In a first approximation (neglecting logarithmic effects) these matrix elements scale like $1\TeV^2/M^2$. Thus after inserting the explicit expressions for the couplings $\Gamma$, they can be easily applied to set bounds on LQ models.

\begin{table}
\centering
\setlength{\tabcolsep}{12pt}
\renewcommand{\arraystretch}{1.5}
\begin{tabular}{@{}ccccc@{}}
\toprule
&$K^0-\bar K^0$ & $D^0-\bar D^0$ & $B_d-\bar B_d$ & $B_s-\bar B_s$
\\
\midrule
$\av{\cO^{(\prime)}_1(\mu)}$ & $0.00202(0.00007)$ &  $0.079(0.003)$ & $0.611(0.058) $ & $0.967(0.068) $
\\
$\av{\cO^{(\prime)}_2(\mu)}$ & $-0.0361(0.0024)$ &  $-0.150(0.010)$ & $-0.508(0.051)$ & $-0.813(0.061)$
\\
$\av{\cO^{(\prime)}_3(\mu)}$ & $0.0124(0.0008)$ &  $0.044(0.004)$ & $0.142(0.030)$ & $0.222(0.033) $
\\
$\av{\cO_4(\mu)}$ & $0.0739(0.0048)$ & $0.252(0.018)$ & $0.921(0.079)$ & $1.367(0.092)$
\\
$\av{\cO_5(\mu)}$ & $0.0154(0.0013)$ & $0.089(0.007)$ & $0.498(0.052) $ & $0.739(0.059)$
\\
\bottomrule
\end{tabular}
\caption{Values for the matrix elements of $\Delta F=2$ operators, in units of $\GeV^4$.
The renormalization scale is $\mu = \{ 3, 3, 4.18, 4.18 \}\GeV$  for $P=\{ K, D, B_d, B_s \}$.}
\label{tab:matrixelements}
\end{table}

\subsection{Matching scale dependence and importance of NLO corrections}

The renormalization-scale dependence of the Wilson coefficients is given by the Renormalization Group Equation (RGE)
\eq{
\Bigg[
\frac{\partial}{\partial \log\mu}
+ \frac{d \alpha_s}{d \log\mu} \frac{\partial}{\partial \alpha_s}
+ \frac{d M_a}{d\log\mu} \frac{\partial}{\partial M_a}
+ \frac{d \Gamma_{q\ell}^{aX}}{d\log\mu} \frac{\partial}{\partial \Gamma_{q\ell}^{aX}}
- \gamma^T
\bigg] \,\vec{C}(\mu) = 0\ ,
}
where a sum over the indices $a,q,\ell$ and $X=L,R$ is understood.
It is easy to check that the matching conditions given in~Eqs.\,(\ref{LQC1})-(\ref{LQC4}) and~(\ref{NLOC1})-(\ref{NLOC5}) satisfy this RGE up to higher order $\cO(\alpha_s^2)$ terms.
More explicitly, using the beta functions in~\Eq{betafcts},
\eq{
\frac{\partial C_i^{(1)}}{\partial \log\mu} = \hat \alpha_s
\bigg[
3C_F M_a \frac{\partial C_i^{(0)}}{\partial M_a}
+ 3C_F \Gamma_{q\ell}^{aX} \frac{\partial C_i^{(0)}}{\partial \Gamma_{q\ell}^{aX}}
+ \gamma^{(0)}_{ji} C_j^{(0)}
\bigg]\ .
}
As discussed already in the previous section, even though the LO Wilson coefficients $C_i^{(0)}$ do not depend explicitly on the matching scale~$\mu_0$, they do depend on it implicitly through the running masses and couplings.
This means we treat $M(\mu_0)$ and $\Gamma_{ij}^X(\mu_0)$ as functions of the matching scale $\mu_0$. Then, one can calculate the matrix element for the $\Delta F=2$ process in question, taking into account that $\mu_0$ is at the same time the initial scale for the renormalization-group evolution of the Wilson coefficients down to the hadronic scale.
The running of the masses and couplings cancels the matching-scale dependence of physical observables order-by-order in $\alpha_s(\mu_0)$.
For the evolution down to the hadronic scale we will use the NLO anomalous dimensions also for the LO estimate (even though this is higher order $\alpha_s$) since these results were known previously to our calculation.

In order to illustrate both the relative size of the NLO matching corrections and the reduction of the matching-scale dependence of physical observables, we focus on the case of $\bar B_s - B_s$ mixing and consider the quantity
\eq{
\cR(\mu_0) = \frac{C_i(\mu_b)\av{\cO_i(\mu_b)}}{C^{(0)}_i(\mu_b)\av{\cO_i(\mu_b)}|_{\mu_0 = 1\TeV}}\ .
\label{Rratio}
}
The numerator in $\cR$ depends on the matching scale $\mu_0$ via the starting scale of the RGE, the LQ mass $M(\mu_0)$, the LQ couplings to fermions $\Gamma^{L,R}(\mu_0)$ and the explicit $\mu_0$ dependence of $C_i^{(1)}$, which contains a logarithm of the matching scale. In the denominator the matching scale is fixed to the reference value $\mu_0=1\TeV$.

\begin{figure}
	\begin{center}
\includegraphics[width=13cm]{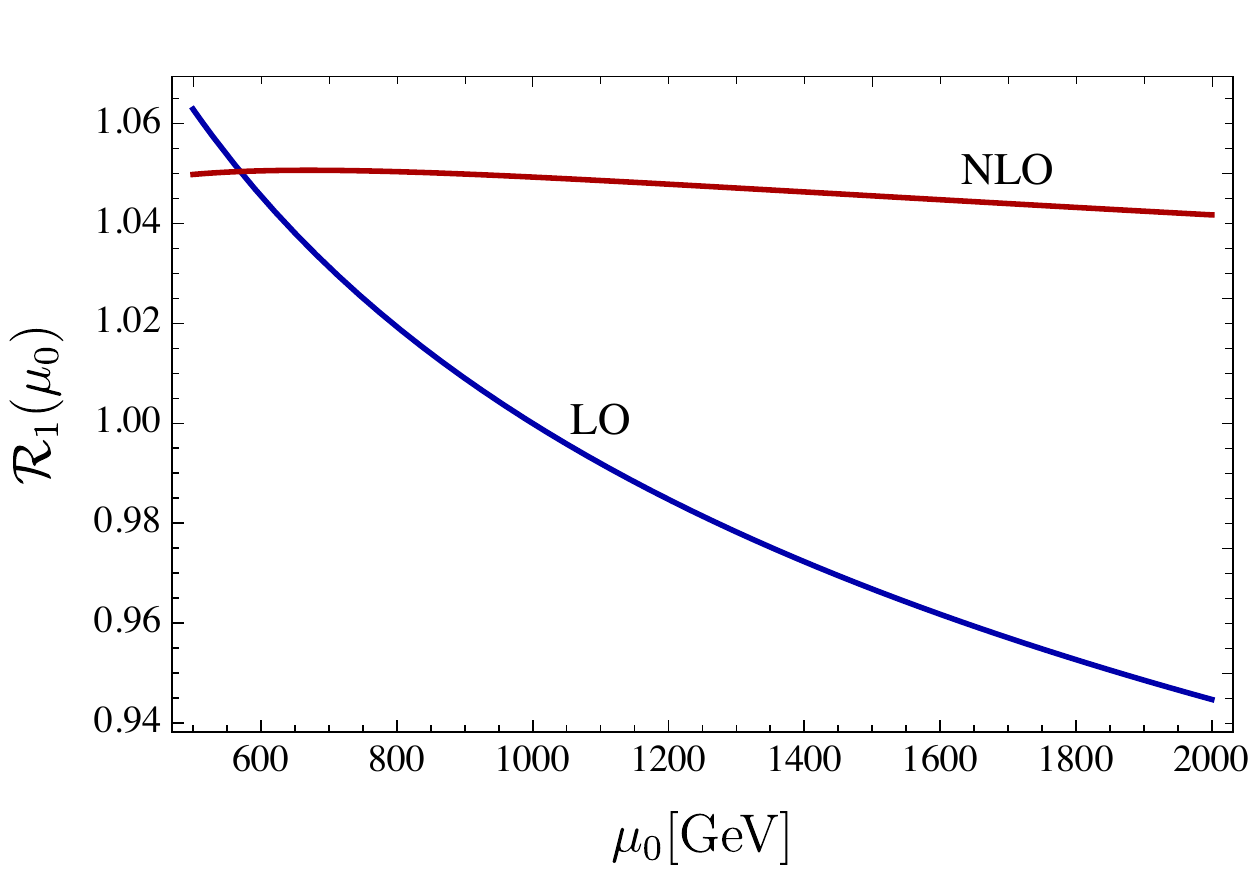}
\hspace{5mm}
\includegraphics[width=13cm]{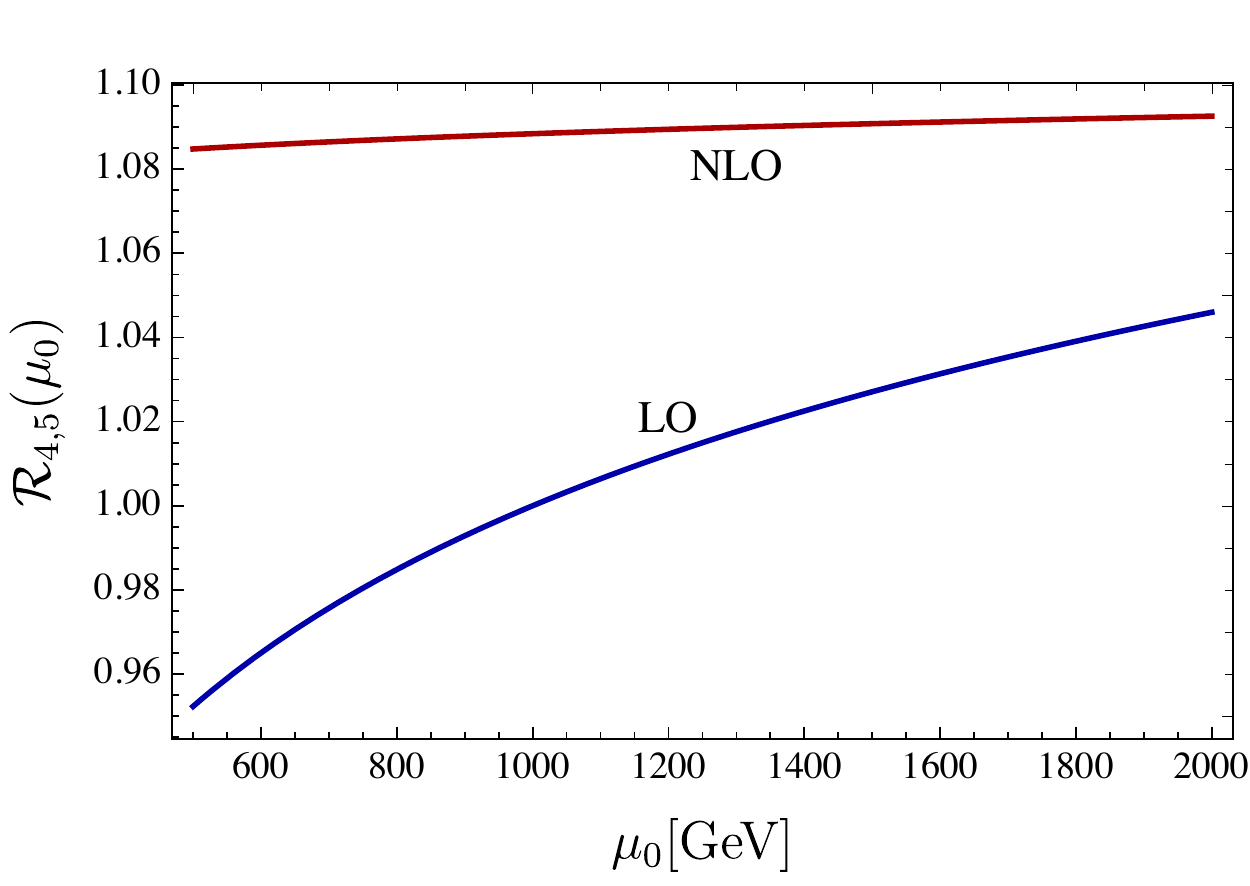}
	\end{center}
\caption{Matching-scale dependence of the ratios $\cR_1$ and $\cR_{4,5}$ for $M(1\TeV)=1\TeV$. The upper plot shows the case in which $C_1$ is generated (the $\Gamma^R\Gamma^R$ contribution to the mass difference), while the lower plot shows the case in which $C_4$ and $C_5$ are generated (the $\Gamma^L\Gamma^R$ contribution to the mass difference).}
\label{C4C5}
\end{figure}

In Fig.~\ref{C4C5} we plot separately the contributions to $\cR$ proportional to $\Gamma^R\Gamma^R$ and $\Gamma^R\Gamma^L$, which are called $\cR_1$ and $\cR_{4,5}$, respectively, as they are related to the corresponding Wilson coefficients. We also show separately the LO and NLO contributions to $\cR$. The LO effect is obtained by setting $C_i = C_i^{(0)}$ in the numerator of~\Eq{Rratio}, understanding $M$ and $\Gamma^{L,R}$ in the expression for $ C_i^{(0)}$ as running parameters at the scale $\mu_0$ derived from their reference values $M(1\TeV)=1\TeV$ and $\Gamma^{L,R}(1\TeV)$. We see that the LO result has a sizable matching scale dependence, both in the  $C_1$ contribution (or equivalently  $C'_1$) and in the $C_{4,5}$ contribution.  
This scale dependence is, as expected and required, significantly reduced once the NLO matching effects are included. One can also see from Fig.~\ref{C4C5} that the NLO corrections lead to a constructive effect of the order of 5\% (8\%) for the case of $C_1$ ($C_{4,5}$).

\section{Conclusions}

Leptoquarks are prime candidates for an explanation of the flavour anomalies, in particular of the hints for NP in $b\to s\ell^+\ell^-$ and $b\to c\tau\nu$ transitions. While contributing to these processes, other flavour observables, such as $\Delta F=2$ processes, are unavoidably also modified. Thus, viable explanations of the flavour anomalies must also satisfy these experimental bounds.

In this article we have studied the $\Delta F=2$ processes $D^0-\bar D^0$, $K^0-\bar K^0$ and $B_{s,d}-\bar B_{s,d}$ mixing in models with scalar LQs. We have first obtained the complete LO matching for all five representations of scalar LQs under the SM gauge group (including mixed contributions). Then we have calculated the NLO $\alpha_s$ corrections to the matching. This allows for a consistent use of the existing two-loop anomalous dimensions in the EFT and significantly reduces the matching scale uncertainty in the calculation of physical observables.  We find that the NLO matching corrections lead to a constructive effect of the order of 5\% (8\%) for the case of $C_1$ ($C_4$, $C_5$).
We have also provided easy-to-use semi-numerical formulas for the neutral meson mixing amplitudes (at the meson level).

\section*{Acknowledgments}
We thank Luc Schnell for checking the results of the 1-loop matching for the five scalar LQ representations. The work of A.C. is supported by a Professorship Grant (PP00P2\_176884) of the Swiss National Science Foundation. A.C. also thanks CERN for the support via the Scientific Associate program. J.V. acknowledges funding from the Spanish MICINN through the ``Ram\'on y Cajal'' program RYC-2017-21870,
the ``Unit of Excellence Mar\'ia de Maeztu 2020-2023” award to the Institute of Cosmos Sciences (CEX2019-000918-M) and from PID2019-105614GB-C21 and 2017-SGR-929 grants.

\appendix

\section{Projections}
\label{app:projectors}

We describe the method used in the evaluation of both the EFT and full theory amplitudes in our calculation. As explained in Sec.~\ref{32section} we construct projectors $P_i, i=1,...,8$ defined as 
\eq{
\label{projectors}
P_i \av{\cO_j}^{(0)} = \delta_{ij} + \cO(\epsilon^2)\,,
}
for a consistent method up to order \(\mathcal{O(\epsilon)}\) as this is the order of the divergences (see Sec.~\ref{correctionsNLO}).
In order to illustrate the approach, we first consider the full theory amplitude. A generic two-loop amplitude can be expressed as 
\eq{
\label{amplitude}
D=\int\frac{d^nk}{(2\pi)^n}\int\frac{d^nq}{(2\pi)^n}\frac{\bar{u}_{s\delta}\Gamma_A(k,q,\mu,\nu)u_{b\alpha}\otimes\bar{v}_{s\beta}\Gamma_B(k,q,\mu,\nu)v_{b\gamma}}{(k^2-m_1^2)^{n_1}(q^2-m_2^2)^{n_2}((k-q)^2-m_3^2)^{n_3}}\,,
}
where Greek letters are colour indices and the $s(b)$ index indicates the \(s(b)\) quark. The objects $\Gamma_{A,B}$ represent strings of gamma matrices and loop momenta with saturated Lorentz indices\,\footnote{
External quark legs are taken massless and with zero momenta.
}. Once the integration is performed (following~\Reff{Davydychev:1992mt}) the amplitude is expressed in terms of strings of gamma matrices with structures \(\overline{\Lambda}^{(i)}_A\otimes\overline{\Lambda}^{(i)}_B\) together with coefficients \(a_i(m)\) which only depend on masses (generically denoted by $m$)
\eq{
\label{2loopamplitude}
D=\sum_{i}a_i(m)\left[\bar{u}_{s\delta}\overline{\Lambda}^{(i)}_Au_{b\alpha}\otimes\bar{v}_{s\beta}\overline{\Lambda}^{(i)}_Bv_{b\gamma}\right].
}
The coefficients $a_i(m)$ are UV finite because the full theory has been renormalized.
The Dirac structure $\bar{u}_{s\delta}\overline{\Lambda}^{(i)}_Au_{b\alpha}\otimes\bar{v}_{s\beta}\overline{\Lambda}^{(i)}_Bv_{b\gamma}$ corresponds to the tree-level matrix element of a physical operator $\langle\mathcal{O}_i\rangle^{(0)}$,
which is always possible after some four-dimensional Dirac algebra.

Therefore, the projectors are constructed such that the coefficients $a_i(m)$ are projected out whenever $P_i$ is applied to the amplitude in~\Eq{2loopamplitude}. For $P_i=P^{(i)}_A\otimes P^{(i)}_B$, we require
\eq{
\label{condition}
\text{Tr}\left[\overline{\Lambda}^{(i)}_AP_A^{(j)}\overline{\Lambda}^{(i)}_BP_B^{(j)}\right]=\delta_{ij}\ ,
}
and the projection on the amplitude is defined by the following replacement:
\eq{
u_{b\alpha}\bar{v}_{s\beta}\rightarrow P_{B}^{(i)} \delta_{\alpha\beta},\quad v_{b\gamma}\bar{u}_{s\delta}\rightarrow P_{A}^{(i)} \delta_{\delta\gamma}\ .
}
Using ~\Eq{condition} we find the projectors
\begin{align}
P_1 &= -\frac{i}{384 }p_1\,, &
P_2 &=  -\frac{i}{64 }p_2+\frac{i}{2304 }p_5 \,,  &
P_3 &= \frac{i}{192 }p_2 +\frac{5i }{2304 }p_5\,,
\nonumber\\
P_4 &= -\frac{i}{192 }p_3-\frac{i}{32 }p_4\,, &
P_5 &= \frac{i}{64 }p_3+\frac{i}{96 }p_4\,, &
\label{4dprojectors}
\end{align}
which are given as linear combinations of the following more basic ones,
\begin{align}
p_1 &= \gamma_{\mu}P_R\otimes\gamma^{\mu}P_R\,, &
p_2 &=  P_L\otimes P_L\,,  &
p_3 &= \gamma_{\mu}P_L\otimes\gamma^{\mu}P_R\,,
\nonumber\\
p_4 &= P_L\otimes P_R\,, &
p_5 &= \sigma_{\mu\nu}P_L\otimes\sigma^{\mu\nu}P_L\,. &
\label{4dprojectors}
\end{align}
The projectors $P'_{1,2,3}$ projecting onto $C'_{1,2,3}$ are equal to $P_{1,2,3}$ with the replacement $P_R\leftrightarrow P_L$.
One can verify that the projectors of~\Eq{4dprojectors} fulfill the condition in~\Eq{projectors}, and so when one applies $P_i$ to an amplitude it projects out the coefficient of the $\langle\cO_i\rangle^{(0)}$ term.

In the EFT, the Dirac algebra cannot be carried out in four dimensions because the presence of $1/\epsilon$ poles in the loop integrals requires keeping $\epsilon$ terms in the Dirac Algebra  (see e.g.~\Reff{Buras:2000if}). Thus, when reducing the spinor structures in the EFT amplitudes to tree-level matrix elements one is forced to introduce Evanescent operators $E=E_A\otimes E_B$.
For the basis of Evanescent operators relevant for $\Delta F=2$ processes we use the ones given in App.~A of~\Reff{Buras:2000if}.
In this case we require the projectors to project out these terms:
\eq{
\label{conditionevanescent}
\text{Tr}\left[E^{(i)}_AP_A^{(j)}E^{(i)}_BP_B^{(j)}\right]=\mathcal{O}(\epsilon^2)\quad\forall\quad i,j\,,
}
This implies that for the EFT the projectors must be calculated to order $\cO(\epsilon)$ explicitly\footnote{The traces with $\gamma_5$ are performed in $D$ dimensions in the Larin scheme.}, and as a consequence the list of projectors \(p_i\) includes further Dirac structures and colour structures. Now, the replacement is
\eq{
u_{b\alpha}\bar{v}_{s\beta}\rightarrow P_{B}^{(k)} S_{\alpha\beta}\,,\quad v_{b\gamma}\bar{u}_{s\delta}\rightarrow P_{A}^{(k)} S_{\delta\gamma}\,,
}
where $S_{\alpha\beta(\delta\gamma)}$ is the colour structure which can either be $\delta_{\alpha\beta(\delta\gamma)}$ or $T_{\alpha\beta(\delta\gamma)}^{A}$. Using the corresponding conditions for the projectors, we find
\eqa{
P_1 &=& -\frac{455i}{2304 }p_1-\frac{193i }{384 }p_9+\frac{i}{1152 }p_8+\frac{323i}{12288 }p_{13}+\epsilon\frac{149i }{256 }p_1\,,
\\
P_2 &=& -\frac{2520983i}{273888 }p_2 +\frac{1489i}{45648 }p_7-\frac{655699i }{45648 }p_{10}+\frac{10229i }{365184 }p_{15}+\frac{7i }{4608 }p_{16}\,
\\
&&-\frac{105617i }{1095552 }p_5+\epsilon  \left(\frac{31272625i }{1643328 }p_2+\frac{4117901i }{136944 }p_{10}\right)\,,
\\
P_3 &=& -\frac{1257571i }{136944 }p_2+\frac{373i }{11412 }p_7-\frac{655645i }{45648 }p_{10}+\frac{10157i}{365184 }p_{15}+\frac{7i }{4608 }p_{16}
\\
&&-\frac{51817i}{547776 }p_5+\epsilon  \left(\frac{15625897i }{821664 }p_2+\frac{2056349i}{68472 }p_{10}\right)\,,
\\
P_4 &=& -\frac{i}{16 }p_{11}+\frac{i}{144 }p_6+\frac{i}{16 }p_{12}-\frac{i}{192 }p_{14}-\frac{i}{12 }p_4-\epsilon  \left(\frac{3i }{16 }p_{12}+\frac{i}{24 }p_3\right)\,,
\\
P_5 &=& -\frac{3i }{16 }p_{12}+\frac{i}{48 }p_{11}-\frac{i}{36 }p_3+\frac{i}{64 }p_{14}+\frac{i\epsilon}{16 }p_{12}\,,
}
with the following list of projectors $p_i$:
For $S_{\alpha\beta(\delta\gamma)}=\delta_{\alpha\beta(\delta\gamma)}$,
\eqa{
p_{6} &=& \gamma_{\mu}\gamma_{\nu}\gamma_{\rho}P_R\otimes\gamma^{\mu}\gamma^{\nu}\gamma^{\rho}P_R\,,
\nonumber\\
p_{7} &=& \sigma_{\mu\nu}P_L\otimes\sigma^{\mu\nu}P_R\,,
\\
p_{8} &=& \gamma_{\mu}\gamma_{\nu}\gamma_{\rho}\gamma_{\sigma}\gamma_{\lambda}P_R\otimes\gamma^{\mu}\gamma^{\nu}\gamma^{\rho}\gamma^{\sigma}\gamma^{\lambda}P_R\,,
\nonumber
}
and for $S_{\alpha\beta(\delta\gamma)}=T_{\alpha\beta(\delta\gamma)}^{A}$,
\begin{align}
p_{9} &= \gamma_{\mu}P_R\otimes\gamma^{\mu}P_R\,, &
p_{13} &= \gamma_{\mu}\gamma_{\nu}\gamma_{\rho}P_R\otimes\gamma^{\mu}\gamma^{\nu}\gamma^{\rho}P_R\,,
\nonumber\\
p_{10} &= P_L\otimes P_L\,, &
p_{14} &= \gamma_{\mu}\gamma_{\nu}\gamma_{\rho}P_L\otimes\gamma^{\mu}\gamma^{\nu}\gamma^{\rho}P_R\,,
\nonumber\\
p_{11} &= \gamma_{\mu}P_L\otimes\gamma^{\mu}P_R\,, &
p_{15} &= \sigma_{\mu\nu}P_L\otimes\sigma^{\mu\nu}P_R\,,
\\
p_{12} &= P_L\otimes P_R\,, &
p_{16} &= \gamma_{\mu}\gamma_{\nu}\gamma_{\rho}\gamma_{\sigma}\gamma_{\lambda}\gamma_{\delta}P_L\otimes\gamma^{\mu}\gamma^{\nu}\gamma^{\rho}\gamma^{\sigma}\gamma^{\lambda}\gamma^{\delta}P_L\,.
\nonumber
\end{align}

\section{Two-loop ADM in the SUSY basis}
\label{app:basis}

Our NLO matching calculation has been performed in the SUSY basis. A byproduct of this calculation is the LO ADM $\gamma^{(0)}$ (which is scheme-independent). However, in order to perform NLL resummation, the NLO ADM $\gamma^{(1)}$ is needed, which arises from the renormalization of the EFT at two-loop order. This calculation has been performed in~\Reff{Buras:2000if}, in a different basis (the ``BMU" basis) for physical operators,
\eq{
Q = \left(Q_1^\text{VLL},Q_1^\text{SLL},Q_2^\text{SLL},Q_1^\text{LR},Q_2^\text{LR},Q_1^\text{VRR},Q_1^\text{SRR},Q_2^\text{SRR}\right)^T\ .
}
Therefore, we must rotate the result in~\Reff{Buras:2000if} to our basis.
The SUSY ($\cO$) and BMU ($Q$) bases are related in the following way
\eq{
\label{BMUtoSUSY}
\vec{\cO}=R\,\left(\vec{Q} + W\vec{E}\right)\ ,
}
where the Fierz-evanescent operators
$\vec{E}^T = \left(\star, E_1^\text{SLL},\star,E_1^\text{LR},\star\right)$ are defined in~\Reff{Buras:2000if}  and
\eq{
R = \left(
\begin{array}{ccccc}
1 & 0 & 0 & 0 & 0 \\
0 & 1 & 0 & 0 & 0 \\
0 & -1/2 & 1/8 & 0 & 0 \\
0 & 0 & 0 & 0 & 1  \\
0 & 0 & 0 & -1/2 & 0
\end{array}
\right),
 \quad W = \left(
\begin{array}{rrrrr}
0 \quad & \quad 0  &\quad 0 &\quad 0 &\quad 0 \\
0 & 0 & 0 & 0 & 0 \\
0 & 8 & 0 & 0 & 0 \\
0 & 0 & 0 & -2 & 0  \\
0 & 0 & 0 & 0 & 0
\end{array}
\right).
}
(The corresponding results for the sector $\cO'_{1-3}$ vs VRR/SRR are the same as the 1-3 sector above.)
 Due to the presence of evanescent operators ($\vec{E}$) and their mixing with the physical operators, the transformation for \(\gamma^{(1)}\) from the BMU basis to the SUSY basis corresponds to a rotation plus a change of scheme, given by (e.g.\cite{Chetyrkin:1997gb, Gorbahn:2004my})
\begin{equation}
\label{transformationADMNLO}
    \gamma^{(1)}_{\mathcal{O}}=R\left(\gamma^{(1)}_{Q}-\comm{r_{Q}}{\gamma^{(0)}_{Q}}-2\beta_0r_{Q}\right)R^{-1}+\comm{r_{\mathcal{O}}}{\gamma^{(0)}_{\mathcal{O}}}+2\beta_0 r_{\mathcal{O}},
\end{equation}
where the matrices $r_\cO$ and $r_Q$ are defined in~\Eq{EFTamp} in their corresponding bases. The LO ADM $\gamma^{(0)}$ is scheme-independent and thus it holds that $\gamma^{(0)}_\cO = R\gamma^{(0)}_Q R^{-1}$.

We compute $r_{Q}$ from scratch directly in the BMU basis (along the lines of~\Sec{setup}), and obtain
\eq{
r_{Q} = \left(
\begin{array}{ccccc}
4/3 & 0 & 0 & 0 & 0 \\
0 & {\color{gray}46/3} & {\color{gray}-1/6} & 0 & 0 \\
0 & {\color{gray}40} & {\color{gray}-6} & 0 & 0 \\
0 & 0 & 0 & 10/3 & -12  \\
0 & 0 & 0 & 0 & 64/3
\end{array}
\right) \log(\mu/\lambda)
+
\left(
\begin{array}{ccccc}
-5 & 0 & 0 & 0 & 0 \\
0 & {\color{gray}5/6} & {\color{gray}-1/8} & 0 & 0 \\
0 & {\color{gray}-46/3}  & {\color{gray}1/6} & 0 & 0 \\
0 & 0 & 0 & -7/6 & 1  \\
0 & 0 & 0 & 3/2 & 19/3
\end{array}
\right)
\ .
}
where again we have indicated {\color{gray}in gray} the sector that does not impact our $SU(2)$-invariant model.
Using~\Eq{transformationADMNLO} and taking $\gamma^{(1)}_Q$ from \Reff{Buras:2000if}, we find

\eq{
\label{NLOADMSUSY}
\gamma^{(1)}_{\mathcal{O}}=
\left(
\begin{array}{ccccc}
\frac{4 f}{9}-7 & 0 & 0 & 0 & 0 \\
0 & {\color{gray}\frac{220 f}{27}-\frac{476}{3}} & {\color{gray}-\frac{4 f}{27}-\frac{4}{3}} & 0 & 0 \\
0 & {\color{gray}73+\frac{110 f}{27}} & {\color{gray}\frac{359}{3}-\frac{218 f}{27}} & 0 & 0 \\
0 & 0 & 0 & \frac{68 f}{9}-\frac{1343}{6} & 4 f-\frac{225}{2} \\
0 & 0 & 0 & \frac{22 f}{3}-99 & \frac{71}{3}-\frac{22 f}{9} \\
\end{array}
\right)\ ,
}
in accordance with~\Eq{gamma1}.

The SLL/SRR sector (in gray) can be reproduced using the formulas given in Section C.1 of~\Reff{Gorbahn:2009pp}, leading to a result in agreement with~\Eq{NLOADMSUSY}.


%
%
%
%
%


\end{document}